\documentclass[a4paper,fleqn]{cas-sc}
\usepackage[authoryear]{natbib}
\def\aap{Astronomy \& Astrophysics}
\def\apj{Astrophyscal J.}
\def\icarus{Icarus}
\def\mnras{Monthly Notices R.A.S}
\def\gca{Geochimica et Cosmochimica Acta}
\def\apjs{Astrophysical J. Suppl.}
\def\apjl{Astrophysical J. Let.}
\def\nat{Nature}
\def\ssr{Space Science Rev.}
\def\maps{Moon and Planetary Sci.}

\begin{document}

\let\WriteBookmarks\relax
\def\floatpagepagefraction{1}
\def\textpagefraction{.001}
\shorttitle{Parent body of EL chondrites}
\shortauthors{M. Trieloff et~al.}

\title[mode = title]{Evolution of the parent body of enstatite (EL) chondrites}

\author[1]{Mario Trieloff}[orcid=0000-0002-2633-0433]
\cormark[1]
%\ead{mario.trieloff@geow.uni-heidelberg.de}
\credit{ Calculations and evaluations for the closure temperature of the I-Xe system in enstatite chondrites, evaluation of the chronological data in order to update age calibrations concerning age monitors and decay constants. Conceptualisation of the study. Interpretation of model fits to thermochronological data, writing of the paper}

\author[1]{Jens Hopp}[orcid=0000-0002-4124-5392]
%\ead{Jens.Hopp@geow.uni-heidelberg.de}
\credit{Calculations and evaluations for the closure  temperature of the I-Xe system in enstatite chondrites, evaluation of  the chronological data in order to update age calibrations concerning age monitors and decay constants. Interpretation of model fits to thermochronological data}

\author[2]{Hans-Peter Gail}[orcid=0000-0002-2190-1794]
%\ead{gail@uni-heidelberg.de}
\cormark[2]
\credit{Parent body model calculations including adaptation of the model to physical properties of enstatite chondrites. Conceptualisation of the study. Interpretation of model fits to thermochronological data, writing of the paper}

\address[1]{Institut f\"{u}r Geowissenschaften, Klaus-Tschira-Labor f\"{u}r Kosmochemie, Universit\"{a}t Heidelberg, Im Neuenheimer Feld 236, 69120 Heidelberg, Germany, mario.trieloff@geow.uni-heidelberg.de, Jens.Hopp@geow.uni-heidelberg.de}

\address[2]{Zentrum f\"{u}r Astronomie, Institut f\"{u}r Theoretische Astrophysik, Universit\"{a}t Heidelberg, Albert-Ueberle-Str. 2, 69120 Heidelberg, Germany, gail@uni-heidelberg.de}

\cortext[cor1]{Corresponding author}
\cortext[cor2]{Principal corresponding author}

\begin{abstract}
Chondrites stem from undifferentiated asteroidal parent bodies that nevertheless experienced a certain degree of metamorphism after their 
formation in the early solar system. Maximum temperatures of metamorphism depend mainly on formation time and the abundance of the main heating source, which is short-lived $^{26}$Al (half life 720 000yr).

Enstatite chondrites formed under reducing conditions and include many strongly metamorphosed members of petrologic type 6. We model the thermal evolution of the parent body of the low metal enstatite chondrite class (EL). The model takes into account accretion, heating, sintering  and compaction by hot pressing of the initially porous material, temperature dependent heat conductivity, and insulation effects by the remaining regolith layer.  

A fit of key parameters of the parent body comprising formation time, radius, and porosity is achieved by fitting thermal histories of EL6 chondrites (LON 94100, Neuschwanstein, Khairpur, Blithfield, Daniel's Kuil) constrained mainly by I-Xe and Ar-Ar ages and their respective closure temperatures. Viable fits are obtained for parent bodies with 120 -- 210 km radius, formed c. 1.8 -- 2.1 Ma after Ca,Al rich inclusions (CAIs), and an initial porosity of 30\%, relatively independent on initial disk temperatures. Optimised models with parent body formation-times $>1.95$ Ma after CAIs imply central core temperatures below incipient plagioclase silicate melting. 

Thermal histories of the different EL6 chondrites are indistinguishable and so are their burial depths. While the exact layering depth is somewhat model dependent (c. 12--20 km), the thickness of the layer from which all five EL6 chondrites stem is $< 1$ km. Hence, an origin from a quite small asteroidal fragment is possible, particular as most excavation ages inferred from cosmic ray exposure data are compatible with a separation as meter sized meteoroids from a small Apollo asteroid 33 Ma ago.

\end{abstract}


\maketitle

%********************************************************************

\section{Introduction}

The asteroids between Mars and Jupiter are remnants of planetary building blocks of the early solar system. High precision isotopic analyses and age dating of meteoritic fragments of asteroids record a chronology of formation of solids and planetesimals that would otherwise not be accessible \citep[e.g., ][]{Tri06}. The oldest objects in the solar system are calcium-aluminium rich inclusions (CAIs) that formed during a brief epoch 4.567 Ga ago in the solar nebula. Planetesimals now located in the asteroid belt --- sourcing the meteorite flux to Earth --- started forming nearly contemporaneously with CAIs \citep{Kle05}. Early formed planetesimals became differentiated bodies with Fe,Ni cores and silicate mantles or basaltic crusts, as the high abundance of short lived  nuclides previously injected by nearby massive stars into the solar nebula liberated sufficient decay energy to heat and melt km sized bodies \citep{Kle05, Tri09}. $^{26}$Al turned out to be the most relevant heat source.
 
More primitive, undifferentiated planetesimals likely accreted after most of the $^{26}$Al decayed, about 2 Ma after CAIs and later \citep[see, e.g., ][]{Miy82,Ben96,Hev06,Sah07,Sah11,Kle08b,Har10}. These undifferentiated bodies stored primitive structures like chondrules, mm-sized former melt droplets that were produced before parent body accretion in the solar nebula. Formation time and original size of these planetesimals can be inferred from the primordial heating and cooling history that is preserved in the meteorites from different layering depths in such parent bodies. These meteorites cooled distinctly due to variable layering depth, parent body material properties (e.g., initial porosity, degree of sintering, heat conduction), parent body size, temperature of the ambient solar nebula, and location of the parent body.
 
Formation time and original size provide important constraints for formation scenarios of planetesimals by streaming instabilities \citep[e.g., ][]{ You05,Joh07,Joh15a,Cuz08} or pebble accretion \citep[e.g., ][]{Lev15,Vis16,Kla20}
during disk lifetimes of a few Myr \citep{Hai01,Pfa15}.

Previous studies already provided constraints on parent body properties inferred from cooling histories, such as for the H, and L chondrite parent bodies \citep{Har10,Hen12,Hen13,Mon13, Gai15,Gai18,Gai19}, the Aca\-pulcoite-Lodranite parent body \citep{Neu18}, or those of iron meteorites \citep{Kle05}.
 
Here, we model the parent body of enstatite chondrites with low metal content (EL chondrites), which originated under exceptionally reducing conditions, likely closer to the sun than ordinary chondrites or carbonaceous chondrites \citep{EGo17,Zel77}.
The cooling histories of such models are compared to the empirical cooling histories of EL meteorites found by analysis of cooling times and closure temperatures for all EL meteorites for which such data exist for at least two different thermochronological systems. By an optimisation procedure we determine a radius and formation time of a model for the parent body and the primordial burial depths of the meteorites within their parent body which reproduces as close as possible the existing meteoritical data.

The plan of our paper is as follows: Section \ref{SectData} discusses the available  set of thermochronological data for EL chondrites, Section \ref{SectModl} describes the thermal evolution model, and Section \ref{SectOpt} the fitting procedure between model and meteoritic data. Section \ref{SectRes} discusses the results and gives our final conclusions. In appendix \ref{SectMaterial} we attempt to determine the material properties required for constructing a thermal evolution model of a body composed of the kind of multi-mineral granular matter found in EL chondrites. 

%********************************************************************

%---------------------------
\begin{table}
\caption{
Closure times $t_\mathrm{closure}$ and closure temperatures $T_\mathrm{closure}$.
}
\label{TabTheChr} 

\begin{tabular*}{\tblwidth}{LCL@{\hspace{.1cm}}L@{\hspace{.3cm}}L@{\hspace{.1cm}}L@{}}
\toprule
Meteorite      & System & \multicolumn{1}{c}{$t_\mathrm{closure}$} & ref. & \multicolumn{1}{c}{$T_\mathrm{closure}$} & ref. \\
               &        & \multicolumn{1}{c}{[Ma]} & & \multicolumn{1}{c}{[K]} \\
\midrule
LON 94100      & Ar-Ar  & $4529.0\pm15$  & (1) &$550\pm70$   & (11) \\
               & I-Xe   & $4557.9\pm1.0$ & (2) &$1000\pm100$ & (12) \\
Neuschwanstein & Ar-Ar  & $4519.0\pm19$  & (1) &$550\pm70$   & (11) \\
               & I-Xe   & $4558.4\pm1.1$ & (2) &$1000\pm100$ & (12) \\
Khairpur       & I-Xe   & $4558.6\pm1.0$ & (6) &$1000\pm100$ & (12) \\
               & Mn-Cr  & $4560.5\pm1.0$ & (3) &$950-1350$   & (13) \\
               & Rb-Sr  & $4481.0\pm36.$ & (4) &$400\pm100$  & (14) \\
Blithfield     & Ar-Ar  & $4522.0\pm31.$ & (5) &$550\pm70$   & (11) \\
               & I-Xe   & $4558.5\pm1.1$ & (6) &$1000\pm100$ & (12) \\
Daniel's Kuil  & Ar-Ar  & $4530.0\pm20.$ & (7) &$550\pm70$   & (11) \\
               & I-Xe   & $4558.1\pm1.0$ & (6) &$1000\pm100$ & (12) \\
Hvittis        & I-Xe   & $4558.6\pm0.8$ & (6) &$1000\pm100$ & (12) \\
               & Hf-W   & $4557.7\pm1.0$ & (8) &$1070\pm50$  & (15) \\
Zak{\l}odzie   & Ar-Ar  & $4530.0_{-10}^{+20}$ & (9) &$550\pm70$   & (11) \\
               & Mg-Al  & $4561.7\pm0.4$       & (10) &$900\pm100$  & (16) \\
\bottomrule
\end{tabular*}

%\bigskip
\renewcommand{\baselinestretch}{0.85}\scriptsize
\par\bigskip\noindent
\parbox{\hsize}{\small 
{\scriptsize
References:}~\scriptsize
\\
Closure times: (1) \citet{Hop14}, (2) \citet{Hop16}, (3) \citet{Shu04}, (4) \citet{Tor93}, (5) average of \citet{Ken81} and \citet{Bog10}, (6) \citet{Ken88}  \citet{Bus08} obtained an I-Xe age of $4558.1\pm0.7$ Ma, indistinguishable from the \citet{Ken88} value, (7) \citet{Ken81}, (8) \citet{Kle09}, (9) \citet{Bog10}, (10) \citet{Sug08}, 
\\[.1cm]
Closure temperatures: (11) Ar-Ar by \citet{Tri03} and \citet{Pel97},  (12) I-Xe this work, (13) Mn-Cr by \citet{Shu04}, (14) Rb-Sr estimation by \citet{Miy82}, (15) Hf-W by \citet{Che14} and \citet{Kle08}, (16) Mg-Al estimated from \citet{LaT98}. 
\par\smallskip}

\end{table}
%--------------------------

\section{Observational material}

\label{SectData}

Previous model calculations of the thermal evolution of chondritic parent bodies were largely dedicated to ordinary chondrite parent bodies and based on U-Pb-Pb and fission track ages of phosphates \citep[e.g.,][]{Goe94}, Hf-W ages of metal silicate separation \citep[e.g., ][]{Kle09}, and K-Ar (Ar-Ar) ages of chondritic plagioclase feldspar \citep[e.g., ][]{Tri03}. Due to the reduced nature of enstatite chondrites \citep[e.g., ][]{EGo17}, particularly phosphate  minerals are absent, and so are U-Pb age data. $^{182}$Hf-$^{182}$W data were reported by \citet{Lee00} for some enstatite chondrites. However, these did not allow to precisely date metal silicate separation. A Hf-W age for EL6 Hvittis was obtained by \citet{Kle09}. Some $^{53}$Mn-$^{53}$Cr ages were measured by \citet{Shu04}. However, only one data point for an EL chondrite (type 6 Khairpur, see Table \ref{TabTheChr}). Moreover, the closure temperature of the Mn-Cr system in enstatite whole rocks is hardly constrained \citep{Ito06,Gan07}. Furthermore, a single Rb-Sr age exists for Khairpur by \citet{Tor93}.
 
On the other hand, Ar-Ar age data were presented by \citet{Bog10} and \citet{Hop14}. While \citet{Bog10} found ``problematic" ages which partly exceeded the age of the solar system, \citet{Hop14} realised that precise isochrons have to be evaluated in order to correct excess argon, which can be considered as part of trapped gases which are much more abundant in enstatite than in ordinary chondrites \citep{Cra81}. Ages recalculated for updated $^{40}$K decay constants \citep{Sch11,Ren11} are listed in Table~\ref{TabTheChr}, however only Ar-Ar ages of LON94100 and Neuschwanstein are corrected for trapped argon. As closure temperature we may refer to the value of $550$ K  \citep{Pel97,Tri03} calculated for sodic chondritic plagioclase, which occurs in both equilibrated ordinary and enstatite chondrites \citep{Zip10,Wei12}. Although other K bearing phases occur, e.g., djerfisherite, these may not contribute a major portion of the K budget. Nevertheless, in order to account for such minor phases with possibly different closure temperature, we assume a somewhat larger error of the closure temperature of 70 K. 
 
%-----------------------------------
\begin{figure}
\centering
\includegraphics[width=.49\textwidth]{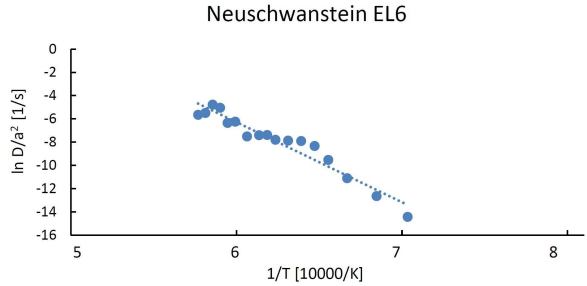}

\caption{Arrhenius plot derived from fractional release of iodine-induced xenon from EL6 chondrite Neuschwanstein. Activation energy $Q$, derived from the slope of the linear regression, and frequency factor $D_0$, derived from the y-intercept, allow to calculate a closure temperature according to the \citet{Dod73} formula.
}

\label{FigArrNeuSw}
\end{figure}
%-----------------------------------

Another chronometer successfully applied to enstatite chondrites is $^{129}$I--$^{129}$Xe dating \citep{Ken88,Hop16}. The situation is quite different to ordinary chondrites, where whole rock I--Xe ages are difficult to interpret due to unknown carrier phases and closure temperatures, and only occasionally feldspar and phosphate mineral separates where dated successfully \citep{Bra99}. In the case of the mineralogically very different enstatite chondrites, early I--Xe studies \citep{Ken81,Ken88} already pointed to enstatite (or inclusions in it) as main carrier phase, a view confirmed also in subsequent studies \citep{Bus08}. The high retentivity of enstatite prevents disturbance by secondary Xe loss and makes I--Xe ages more reliable. Recent calibrations involving the Shallowater standard increased accuracy, corresponding I--Xe ages reported by \citet{Hop16} and \citet{Ken88} --- corrected for recent calibrations of the Bjurb\"{o}le versus the Shallowater standard \citep[see ][]{Hop16} --- are listed in Table~\ref{TabTheChr}. An important parameter not yet constrained, however, is the closure temperature, which is evaluated in the following.
 
Similar to Ar-Ar dating, closure temperatures can be calculated for stepwise degassing experiments, particularly if degassing experiments are performed with a high number of degassing steps. The xenon fraction released at specific temperatures allows us to calculate diffusion coefficients according to equations by \citet{Car59} and \citet{Rei53}, taking into account the temperature dependence of the diffusion coefficient:
\begin{equation}
 D=D_0\,\mathrm{e}^{-Q/R_\mathrm{g}T}\,,
\end{equation}
where $T$ is temperature, $R_\mathrm{g}$ the gas constant, $Q$ the activation energy, and $D_0$ the pre-exponential or frequency factor.
 
Plotting $\ln D/a^2$ versus $1/T$ in arrhenius diagrams allows to calculate $Q$ from the slope and $D_0/a^2$ ($a$: grain radius) from the intercept of a linear regression, and from this a closure temperature from the \citet{Dod73} formula \citep[e.g. ][]{Tri94,Pel97}. In detail, we proceded as follows: According to the procedure as described in \citet{Tri98} and \citep{Pel97}, we excluded the low temperature gas release from calculation of $\ln D/a^2$, as low temperature reservoirs seriously disturb the results, i.e. by shifting  slope and activation energy to erroneously low values. For the  calculations we used the neutron induced iodogenic $^{128}$Xe fractional release of high temperature extractions, from which chronological data were inferred. From the samples analysed we used only linear arrays with correlation coefficients better than 0.9 (Neuschwanstein, see Fig. \ref{FigArrNeuSw}, LON94100, Khairpur) which yielded activation energies of 132--137 kcal/mol and $\ln D_0/a^2$ values of 31--35, and corresponding closure temperatures of c. 990 K. Considerung possible systematic uncertainties, we assign a preliminary estimate of the I-Xe closure temperature of $1\,000\pm100$ K. We note that this value is indistinguishable from the estimate by \citet{Hop16} of $1\,020\pm100$ K, which was, however, a more rough estimate based on laboratory degassing temperatures of xenon versus argon.

In Table \ref{TabTheChr} we list the EL chondrites which were successfully dated by at least two thermochronological tools (I-Xe, Ar-Ar, Mn-Cr, Rb-Sr). All these samples belong to petrologic type 6, which is the dominant petrologic type among EL6 chondrites. EL6 chondrites are reported to have low shock stage S2 \citep{Rub97}, which is important to avoid (or at least minimise) secondary disturbance of chronometers by impact metamorphism. For the optimisation procedure (see below), we excluded Hvittis, as the two data points (Hf-W and I-Xe age) are essentially indistinguishable, nevertheless, we note the good agreement of the Hvittis I-Xe age with other EL6 samples. Furthermore, we also excluded Zak{\l}odzie, as its genesis may involve impact metamorphism (see discussion below), and we exclude the Mn-Cr data of Khairpur because of the highly uncertain closure temperature.

%********************************************************************

\section{Thermal evolution model}

\label{SectModl}

%-------------------------------------------------------------------------------
\subsection{Model assumptions}

A model of the thermal evolution is calculated for a body with asteroidal size and material composition supposed to be typical for EL chondrites. The basic assumptions of our model calculation are, 
\begin{itemize}
\item first, that the body is formed out of the cold material in the accretion disc of the proto-sun and is internally heated by decay of short-lived radioactives (particularly by $^{26}$Al) and cooled by heat transport to the surface where the excess heat is radiated away, 
\item second, that the body forms and grows to essentially its final radius within a period short compared to the half-life of $^{26}$Al such that it starts its evolution cold before much of the decay energy is liberated,
\item and third, that the material forming the body shows initially an on-the-average homogeneous composition all over the body despite its complex micro-structure. 
\end{itemize}
We assume that during the evolution of the parent body of EL chondrites the peak temperature remains sufficiently low that the mineral and metal components of the chondritic material do not start melting at the burial depth of the EL6 chondrites, and that no significant partial melting occurs in the center of the body, such that no large changes in the chemical and mineralogical composition of the material and no differentiation under the action of gravity need to be considered. 

We consider, however, that the initially chondrule-domi\-nated granular material can be compacted at elevated temperature and pressure within large parts of the body's interior. The compaction by hot pressing needs to be considered since it causes an increase of the heat conduction within the material by elimination of the initially considerable fraction of voids in the granular material. It is also responsible for much of the structural changes of the material during the course of the evolution of the chondritic material from petrologic type three to type six. Other processes responsible for the equilibration of the material during its thermal evolution are of minor importance for the temperature evolution and are not considered.

The local deformations by creep at high temperature, if active, deform at the same time an initially irregular shaped body as a whole into a near-spherical shaped configuration. This allows us to consider for simplicity a spherically symmetric built-up body.

An important point in our model is the question if the duration of the growth process can be neglected or not. According to our present understanding, the asteroids form within the accretion disk of the early proto-sun by agglomeration from tiny, sub-$\mu$m-sized dust grains to asteroidal sized bodies of the order of hundred to a few hundred kilometres radius.  In classical bottom-up accretion models of planetesimals \citep[e.g. ][]{Wei77}, gravity-assisted growth by $>$1 km sized planetesimals is fast. The problem of these --- likely obsolete --- classical models are a number of barriers tested in the laboratory: the bouncing, the break-up, and the drift barrier \citep[e.g., ][]{Mor16}. More recent models circumvent these problems by assuming that planetesimals are formed from clumps of solid particles that reach densities high enough to become self-gravitating and contract to form the final bodies,  a process also favoured by a disk's turbulence  \citep{Joh07,Cuz08} and the Kelvin-Helmholtz instability \citep{You05}. This is the so-called streaming instability \citep{Joh07,Joh15,Joh15a}, and this process is very fast. A further concept, called pebble accretion, is based on pebbles drifting inwards in the protoplanetary disk. These can be captured by planetesimals, causing their size growth to planetary embryos and even further \citep{Lev15,Vis16, Kla20}. During the initial formation stages bodies are too small to be significantly heated by decay of radioactives because the liberated heat is efficiently transported to the surface and transferred to the accretion disk. During this phase the material of the just forming bodies is essentially at the same temperature as the mid-plane temperature of the accretion disk. This is true as long as bodies are less than c. km in size. Bodies of the order of hundred kilometres size, on the contrary, are too big for rapid energy-loss by heat conduction and start to heat up. This means that in our models of thermal evolution, we have to consider the growth in the size regime of significantly bigger than km sized bodies. 

Whether we consider gravity-assisted growth in the $> 1$ km regime, or scenarios based on streaming instabilities or pebble accretion, all these concepts favour a rapid growth within the size regime in which internal heat is retained effectively. The growth process is particularly short compared to the decay time-scale of the most efficient heat supplier, $^{26}$Al (half-life of 0.72 Myr). That is, the essential heating of the interior of such bodies occurs only if they have already achieved almost their final size. Hence, we follow the common practice to apply the instantaneous formation approximation where it is simply assumed that the body is suddenly formed at some birthtime, $t_\mathrm{b}$, with some initial temperature, $T_\mathrm{i}$ (assumed to be constant across the body; heating by release of gravitational energy during growth becomes important only for much bigger sized bodies). The initial temperature, $T_\mathrm{i}$, should equal the mid-plane temperature of the accretion disk at the distance of the birthplace of the body from the proto-sun at the instant $t_\mathrm{b}$. We will also argue below in the discussion, that even if final layers are accreted up to 2 Myr after the birthtime, this will not alter our conclusions, because this corresponds to 4 Myr after CAIs, a time when possible pebble fluxes have mostly ceased \citep{Kla20}.

%-------------------------------------------------------------------------------
\subsection{Equations for thermal evolution}

The raise and fall of temperature within the body is described by the heat conduction equation which in case of spherical symmetry is
\begin{equation}
\varrho c_V\,{\partial\,T\over\partial\,t}=-{1\over r^2}\,{\partial\over\partial\,r}\, r^2\,K\,{\partial\,T\over\partial\,r}+\varrho h\,.
\end{equation}
Here $r$ is the distance from the centre, $\varrho$ the mass density of the material, $c_V$ and $K$ the heat capacity per unit mass and the heat conductivity of the material, respectively, and $h$ is the energy production rate by per unit mass of the material. 

The chondritic material contains initially a significant fraction of voids. The density $\varrho$ then is given by 
\begin{equation}
\varrho=\varrho_\mathrm{b}\left(1-\phi\right)\,,
\end{equation}
where $\varrho_\mathrm{b}$ is the density of the compact, void-free material and the porosity, $\phi$, is the fraction of the volume allotted to the empty space between the mineral and metal grains.

The heat conductivity, $K$, is strongly dependent on $\phi$, typically like 
\begin{equation}
K=K_\mathrm{b}\,f(\phi)\,,
\label{DefKpor}
\end{equation} 
where $K_\mathrm{b}$ is the bulk heat conductivity of the pore-free material ($\phi=0$) and some function $f(\phi)$ which depends on the pore-geometry \citep{Hen16}. This is specified in Sect.~\ref{SectHeatCond}.

The porosity, $\phi$, changes with time under the action of the pressure, $P$, acting on the material. The pressure loading at each radius $r$ inside the body is given by the hydrostatic pressure equation
\begin{equation}
{\partial\,P\over\partial\,r}=-{G\,M(r)\,\varrho\over r^2}\,,
\end{equation}
where $G$ is the gravitational constant and $M(r)$ is the mass contained within a sphere of radius $r$.

The variation of the porosity is described by a differential equation
\begin{equation}
{\partial\,\phi\over\partial\,t}=F\left(\phi,P,T\right)\,.
\end{equation}
The details are specified in Appendix~\ref{SectSint}. The compaction process is strongly temperature dependent. There is practically no compaction below some characteristic temperature $T_\mathrm{cmp}$ while compaction rapidly runs into completion at somewhat higher temperature \citep[cf. Appendix A5 of ][]{Gai19}. Because of this, the initially homogeneously composed body develops a structure with a compacted (undifferentiated) rock-like core outlining the region where the peak temperature during the raise and fall of the temperature at some location inside the body exceeded the temperature $T_\mathrm{cmp}$, and a sandstone-like porous mantle where the peak temperature always stayed below~$T_\mathrm{cmp}$. 

For solving the equations we use instead of the radius variable $r$ (varying between zero and the radius, $R$, of the body) a mass variable $m$ (varying between zero at $r=0$ and the total mass, $M$, of the body at $r=R$). The radius, $r$, is calculated at each instant from
\begin{equation}
{\mathrm{d}\,r\over\mathrm{d}\,m}={1\over4\pi\varrho_\mathrm{b}(1-\phi) r^2}\,.
\end{equation}
This automatically accounts for the shrinking of the body during compaction. The total mass of the body, $M$, is constant, the radius of the body, $R$, varies with time while the material is compacted by the action of self-gravity and elevated temperature. Hence, we have to discriminate between the actual radius, $R$, and the pre-compaction radius, $R_\mathrm{pr}$, at birthtime of the body. We prefer to characterize the body by its initial radius $R_\mathrm{pc}$.  

%-------------------------------------------------------------------------------
\subsection{Initial and surface temperature}

\label{SectIniSrfTem}

It is necessary for the construction of an evolution model to specify the initial temperature and the surface temperature during the period considered.

The initial temperature is not known and cannot uniquely be determined in our case. If sufficient data for the cooling history of meteorites are available, it can be determined from model fitting, as it was done for the case of the H chondrite parent body by \citet{Hen13}. This requires in particular data for chondrites of low petrologic type because their cooling history is strongly affected by the value of the initial temperature. Because we have only data for EL6 chondrites which suffered high peak temperatures and, thus, need to originate from large burial depths, this procedure to fix the initial temperature can presently not be applied for the case of EL chondrites. Instead there is some kind of degeneracy between the formation time of the body and its initial temperature (see Appendix~\ref{SecTempDepMod}) in the sense that an increase in initial temperature  can be compensated by a certain increase of formation time without changing the later thermal evolution. For this reason we assume the initial temperature to be a free parameter that is varied in the range between between 200 K and 400 K. 

%-----------------------------------
\begin{figure}
\centering
\includegraphics[width=0.5\textwidth]{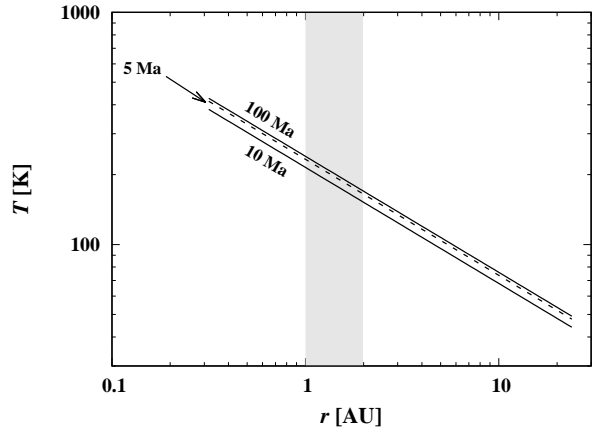}

\caption{Radial stratification of the radiative equilibrium temperature of a body with the albedo of enstatite chondrites at 5, 10 and 100 Myr evolution time of the young sun. The vertical gray bar indicates the distance range where the parent bodies of the enstatite chondrites probably formed.
}

\label{FigTempInBd}
\end{figure}
%-----------------------------------

With respect to the boundary temperature we remark the following: During the first few million of years of evolution of the body after its formation the body is still located within the dusty accretion disk of the proto-sun. Its surface temperature, $T_\mathrm{s}$, is determined during this initial part of its evolution by radiative energy exchange with the disk and by ventilation, and therefore is close to the temperature of the disk (and, thus, also to $T_\mathrm{i}$). After disk dissipation the temperature at the surface of the body is  mainly determined by heating due to illumination by the proto-sun and cooling by radiation to space (the heat flux from the interior turns out to be small compared to the radiative terms and is of negligible influence). This equilibrium temperature is shown in Fig.~\ref{FigTempInBd} at 5, 10, and 100 Myr after formation of the sun, calculated for enstatite chondrite material with an albedo of 0.2 and using an evolution model for a 1 M$_\odot$-mass star from \citet{Dan94}. This temperature is likely to be lower than the temperature $T_\mathrm{s}$ at the surface immediately after formation of the body (equal to the mid-plane temperature of the accretion disk out of which the body formed). It is not really constant since the luminosity of the young sun is not constant, but its variation during the crucial period of the thermal evolution extending from a few Myr to about 100 Myr after formation of the sun is sufficiently small as that it can be treated as constant. Inspection of Fig.~\ref{FigTempInBd} shows that a typical value for this temperature at the distance range where the parent bodies of E chondrites likely formed is 200 K, and this value will be used in our model calculations. 

A formation region at c. 1.7 to 2 AU would be compatible with the radial distance of the Hungaria asteroid family that seems to be spectroscopically linked to the enstatite achondrite class of the aubrites \citep[e.g.,][]{Bin14,Luc19}. These have a reduced mineralogy similar to enstatite chondrites, although they are not derived from the same parent body \citep{Zha96b}. Other studies suggest formation regions closer to the sun, based on isotopic similarities to terrestrial values \citep[e.g.,][]{ Jav10,Dau17}. Hence, we consider a possible formation region between 1 and 2 AU.

The lowering of the surface temperature due to disk dissipation is likely to have some influence on the temperature history of meteorites of low petrologic type because these stem from shallow layers below the surface of the body. For these, the temperatures only suffer a short initial excursion to elevated temperatures followed by rapid cooling within the first few million years which is strongly influenced by the surface temperature. For meteorites of high petrologic types, however, which come from deeper layers where the temperature excursion to temperatures exceeding 1\,000 K lasts for several tens of Myr, the precise variation of the boundary temperature during the first few Myr of thermal evolution is only of minor importance for their thermal history. In the case of EL chondrites no meteorites of low petrologic type suited to study their temperature history are pre\-sently known, only such of type EL6. This allows us to assume for simplicity a constant value for the boundary temperature~$T_\mathrm{s}$ in our models.

%-------------------------------------------------------------------------------
\subsection{Initial porosity}

According to our present knowledge asteroids form by some process that assembles free floating dust and chondrules from the accretion disk into planetesimals \citep[see][for a review on asteroid formation]{Joh14,Joh15,Bir16}. The details of this process are not important for us, only the fact that initially the material is a loosely packed bimodal granular medium consisting of about mm-sized granules and about $\mu$m sized dust. The growth of bodies to the size of the parent bodies of meteorites involves according to presently discussed models low relative velocities of $\lesssim1\,\rm km\,s^{-1}$ which are much too low to shatter dust particles and chondrules. The initial structure of the material from which the bodies are built is therefore assumed to be a granular material which to some extent may be slightly compressed by the action of self-gravitation of the body and by collisions during its growth.  

Information on the initial structure of the material is preserved in the meteorites of petrologic type three. During the course of the chemical evolution of the parent body the initial porosity disappears by sintering. The porosity found in petrologic type six meteorites \citep{Bri03} is of a secondary nature resulting from action of shocks, for instance during the excavation of the meteorites, or reflects weathering, if derived as modal porosity \citep{Bri03}. This does not tell us anything about the initial structure of the material. In the case of enstatite chondrites only very few type 3 chondrites are known. \citet{Wei14} studied a number of EL3 and EH3 chondrites and note that the matrix in two EL3 chondrites has low abundance and that, for instance, MAC 88136 appears to be matrix-free. The dusty rims of chondrules frequently seen in ordinary chondrites seem also to be absent in enstatite chondrites \citep{Krot05}. For EH3 chondrites the matrix abundance also is generally found to be low, though in ALH 81189 a rather high fraction of 14 vol\% is found \citep{Wei14,Rub09}. 
 
In enstatite chondrites we likely have to deal with an initial structure of the material corresponding to the random packing of near-spherical chondrules from a very limited size range and mean diameter of $\sim0.5\dots0.6$ mm \citep{Fri15} with a small or nearly absent additional admixture of $\mu$m-sized dust particles. Without dust and for mono-sized spherical chondrules the porosity of the granular material would be $\phi=0.44$ for the loosest closed packing and $\phi=0.36$ for the random closed packing. Which type of packing is really encountered depends on the way how asteroids assemble from their smallest building blocks and the involved relative speeds, which are not known. An admixture of matrix of up to 15 vol\% would change the porosity to about $\phi=0.45$ to $\phi=0.25$, respectively \citep{Gai15}. In principle, initial porosities between about $\sim0.40$ and $\sim0.25$ could be possible and are considered in our calculations.

We assume in our model computations that initially the porosity is constant throughout the body. This assumption is only relevant, however, for the outer layers of the body since the temperature inside the body reaches after ca. 1 Ma such a level ($\gtrsim1\,000$ K) that the material becomes compacted. By this, any remembrance on the initial structure is lost for that part of the body. 

%********************************************************************

%---------------------------
\begin{table}
\caption{Meteorites and Thermochronological systems used for the optimisation}
\label{TabMetOpt}
\begin{tabular}{@{}L@{\hspace{.5cm}}L@{\hspace{.5cm}}L@{}}
\toprule
Meteorite & \multicolumn{2}{c}{System} \\
\midrule
LON94100       &  Ar-Ar & I-Xe  \\
Neuschwanstein &  Ar-Ar & I-Xe  \\
Khairpur       &  Rb-Sr & I-Xe  \\
Blithfield     &  Ar-Ar & I-Xe  \\
Daniel's Kuil  &  Ar-Ar & I-Xe  \\
\bottomrule
\end{tabular}
\end{table}
%--------------------------

\section{Optimised evolution model}

\label{SectOpt}

The set of thermochronologic data of EL chondrites discussed in Section~\ref{SectData} is used to determine the radius $R$, the birth-time $t_\mathrm{b}$, and other properties of a thermal evolution model which matches the thermochronological data as close as possible. This requires that the parent body underwent a relatively undisturbed evolution without catastrophic collision events with other bodies during the first about 100 Ma evolution of the solar system, such that the onion-shell model hypothesis applies. If this assumption is valid or not can only be judged a posteriori if the thermochronologic data can consistently be interpreted within the frame of this model.

%-----------------------------------
\begin{figure*}
\centering
\includegraphics[width=.49\textwidth]{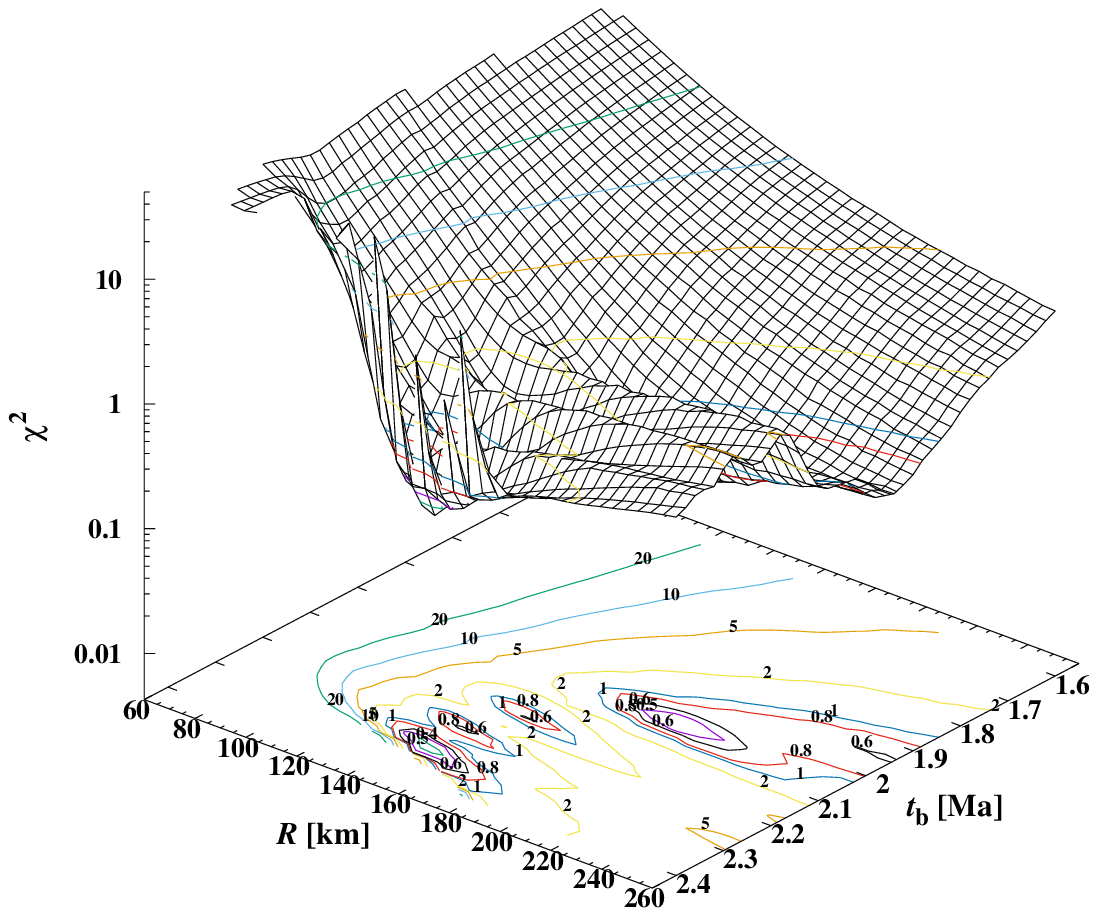}
\includegraphics[width=.49\textwidth]{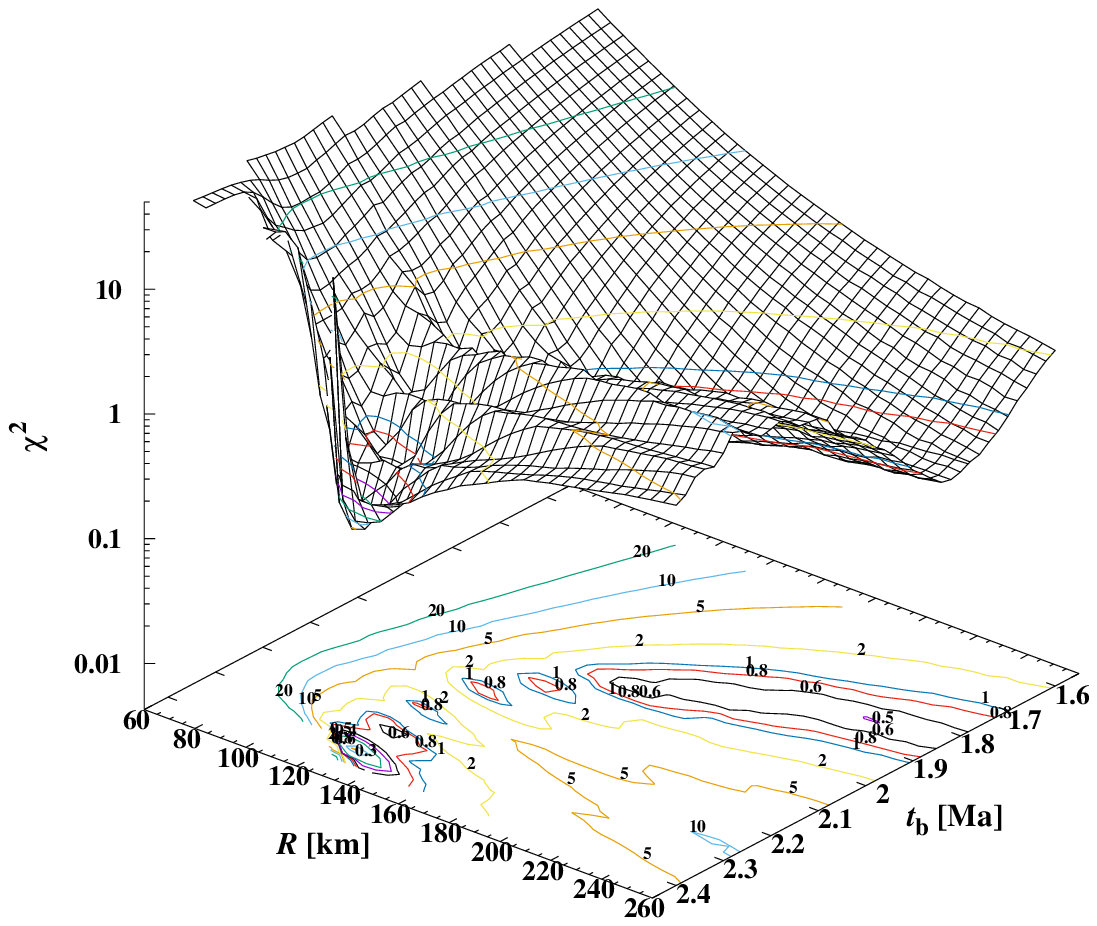}

\includegraphics[width=.49\textwidth]{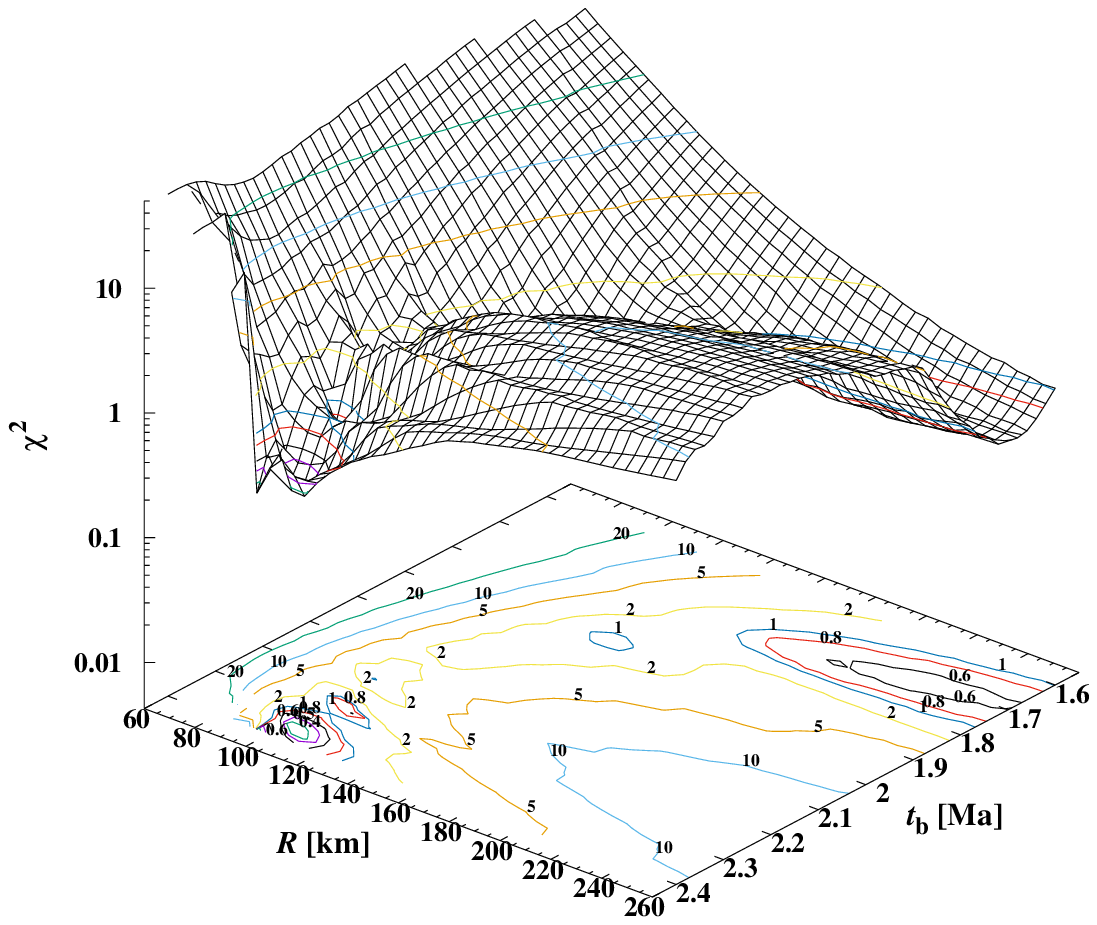}
\includegraphics[width=.49\textwidth]{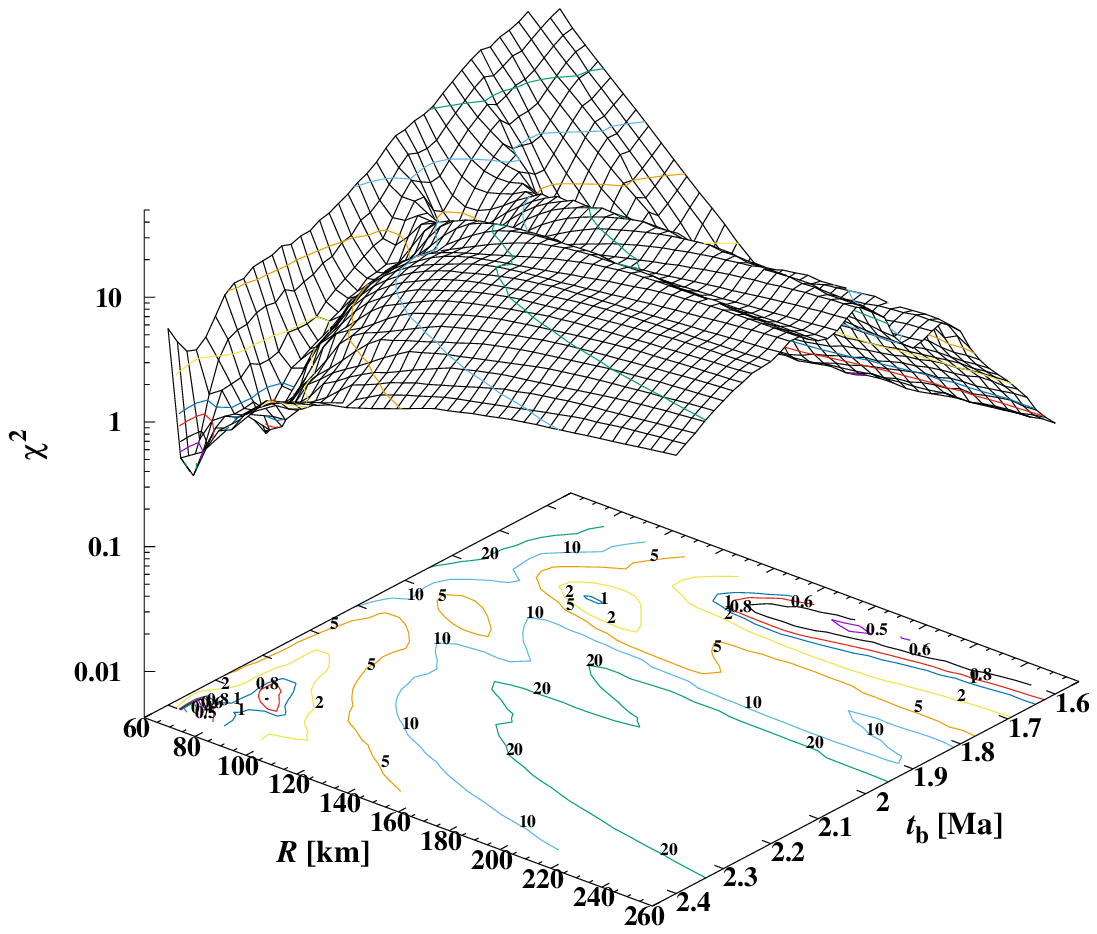}

\caption{
Variation of quality function $\chi^2$ with radius $R$ and formation time $t_\mathrm{b}$ of the body for different initial porosities $\Phi_0=0.25$ (upper left), $\Phi_0=0.30$ (upper right), $\Phi_0=0.35$ (lower left), $\Phi_0=0.40$ (lower right) and initial temperature $T_\mathrm{i}=300$ K. Contourlines for constant values of $\chi^2$ are also shown as projections into the base-plane.
}

\label{FigThermEvo}
\end{figure*}
%-----------------------------------

We selected for our purpose the five meteorites and the two thermochronological sytems for each of them shown in Table~\ref{TabMetOpt}. This data set comprises the meteorites for which sufficiently reliable closure time and closure temperature data are available. The ten different datapoints are opposed by seven unknown quantities: besides the five burial depths of the meteorites we have to determine the radius and birthtime of the body. This leaves us with three more data points than variables. The available dataset is quite scarce but should suffice to find a meaningful fit between data and model.  

%------------------------------------------------------------------------------
\subsection{The quality function}

\label{SectChi}

The method how we obtain a fit is the same as in \citet{Hen13} and \citet{Gai19}. It is done by two inter-nested optimisations. First, for a model with given $R$ and $t_\mathrm{b}$ for each of the chondrites ($n=1,\dots,5$) a burial depth $b_n$ is determined by a least square minimisation. 

For this, we first determine for each of the thermochronological systems (denoted by index $j$) for which we have closure time determinations for this meteorite the temperature $T_{n,j}$, which is the temperature of the model, $T(r,t)$, at some depth $r=b_n$ and at instant $t=t_{\mathrm{closure},j}$. That is the temperature of the model at the measured closure time. We take 
\begin{equation}
\chi_n^2=\sum\limits_{j=1}^2{\left(T_{n,j}-T_{\mathrm{c},n,j}\right)^2\over\sigma^2_{\mathrm{T},n,j}}\,,
\end{equation}
as a measure of how good the temperature variation of the model at depth $b_n$ reproduces the thermochronological data of the meteorite at $t_{\mathrm{closure},j}$. Here $\sigma_{\mathrm{T},n,j}$ are the errors of the closure temperature determination. By minimising $\chi_n^2$ we determine $b_n$ and the corresponing value of $\chi_n^2$ at minimum. 
 
Next, the resulting values of $\chi_n^2$ at minimum are added to obtain our quality function 
\begin{equation}
\chi^2=\sum_n\chi_n^2
\label{EqQual}
\end{equation}
for the final optimisation of the remaining two free parameters, $R$ and $t_\mathrm{b}$. This second optimisation is done by a genetic algorithm.

%------------------------------------------------------------------------------
\subsection{Variation of $\chi^2$ with radius and formation time}

Before turning to the final optimisation, we first calculate $\chi^2$ for a set of models with different radii, $R$, and birthtimes, $t_\mathrm{b}$, and for four values of the initial porosity $\Phi_0$. Figure~\ref{FigThermEvo} shows the variation of the quality function. This is what we obtain in the first step of our optimisation procedure, if for given $R$ and $t_\mathrm{b}$ for each of the meteorites a best-fit burial depth is determined. The range of radii, $R$, is chosen here as the likely range of radii for meteoritic parent bodies from 60 km to 260 km. Bodies smaller than about 60 km radius would cool too rapidly to be compatible with the existence of a period of several million years with temperatures above 1\,000 K as required by the existence of meteorites of petrologic type six. Bodies bigger than about 260 km radius should once have been completely molten and differentiated, which is not suggested by the properties of EL chondrites. The range of formation times considered is chosen from 1.6 Ma to 2.4 Ma after formation of CAIs for analogous reasons.   

The models are calculated for four different values of the initial porosity $\Phi_0$ between the extreme cases $\Phi_0=0.25$ and $\Phi_0=0.40$, because $\Phi_0$ is not well constrained from properties of primitive meteorites. The figure shows, that in all cases the quality function $\chi^2$ takes low values along a curved region in the $R$-$t_\mathrm{b}$ plane. This results from the fact, that of our meteoritic sample four of the five meteorites have data only for I-Xe and Ar-Ar clocks while the fifth has I-Xe and a Rb-Sr closure times, and that the I-Xe closure times of all of the meteorites are almost identical and of high accuracy. These meteorites then must originate from almost identical burial depths within the parent body in order that the temperature at their burial depths passes for all of them at almost the same time through the closure temperature of the I-Xe thermochronometer. On the other hand, the closure times for the Ar-Ar and Rb-Sr thermochronometres have rather large uncertainties such that the requirement with respect to the temperature evolution at the burial depth of the meteorite to pass within the error-limits trough that data point can be met by a variety of $R$-$t_\mathrm{b}$ combinations. The data then do not strongly constrain a single $R$-$t_\mathrm{b}$ combination that allows to match the meteoritic data but merely result in a $R$-$t_\mathrm{b}$-relation for possible radii and birthtimes of the parent bodies.

%-----------------------------------
\begin{figure}
\centering
\includegraphics[width=.49\textwidth]{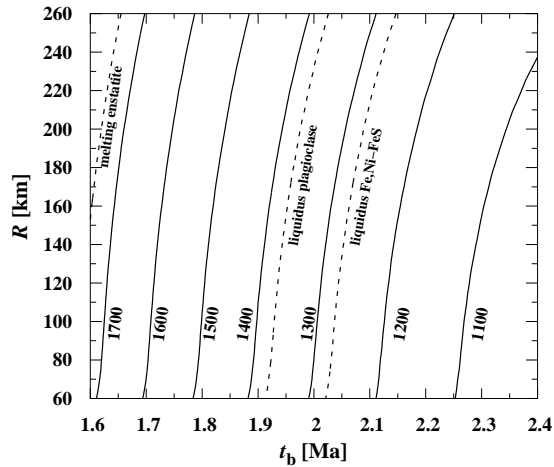}
\caption{
Variation of maximum central temperature during thermal evolution of an asteroid with radius $R$ and formation time $t_\mathrm{b}$ of the body. Dashed lines correspond to the onset of melting of the major components of the enstatite chondrite material  (the tempatures are only rough estimates, see text). An initial temperature of 300 K and initial porosity of $\Phi_0=0.30$ are assumed.
}

\label{FigTc}
\end{figure}
%-----------------------------------

This degeneracy of our data set can be somewhat reduced by giving the data of Khairpur a higher weight in the quality function than for the other four meteorites because the closure temperature of the Rb-Sr clock is different from that of the Ar-Ar clock of the four other meteorites, such that the preponderance of the Ar-Ar clock in the available data-set is somewhat alleviated. In the models in Fig.~\ref{FigThermEvo} we applied a weighting factor of three on the contribution $\chi_n^2$ of Khairpur to~$\chi^2$. 

Even if we proceed in this way, we do not see in  Fig.~\ref{FigThermEvo} one single and well defined minimum in the $\chi^2(R,t_\mathrm{b})$ surface that suggests a well defined radius-birthtime combination for the parent body of the EL chondrites. What we do have are a few rather shallow minima. For the two smaller initial porosities $\Phi_0=0.25$ and $\Phi_0=0.30$ the situation looks slightly more favourable than for the higher values $\Phi_0=0.35$ and $\Phi_0=0.40$.
 
%------------------------------------------------------------------------------
\subsection{Temperature restrictions}

Figure~\ref{FigTc} shows the variation of the maximum of the central temperature attained during the evolution of the body. This has to exceed the metamorphic temperature at which EL6 chondrites are equilibrated, because we have such meteorites at our hands, while, on the other hand side, it has to stay below the limit temperature above which melting of the interior region of the body would occur because there are no indications for large-scale melting of the parent body of the EL chondrites.  

%---------------------------
\begin{table*}[width=1.0\textwidth,pos=h]
\caption{
Best-fit models for the parent body of EL chondrites.
}
\label{TabOptMod}

\begin{tabular*}{\tblwidth}{@{}LLLLLLLL@{}}
\toprule
Quantity & \multicolumn{6}{c}{}   & Unit \\
\midrule
Model                              & (A)   & (B)   & (C)    & (D)   & (E)     & (F)   & \\[.1cm]
            &   &  & \multicolumn{2}{c}{\bf Parameter} &  &  &  \\[.1cm]
Initial temperature $T_\mathrm{i}$ & 250   & 300   & 350   & 300   & 300     & 300   & K  \\
Surface temperature $T_\mathrm{s}$ & 200   & 200   & 200   & 200  & 200     & 200   & K  \\
            &  &  & \multicolumn{2}{c}{\bf Parent body} \\[.1cm]
Initial radius $R_\mathrm{pc}$
                                   & 147   & 139   & 139   & 179   & 232     & 131   & km \\
Final radius $R$                   & 137   & 129   & 127   & 162   & 209     & 122   & km \\
Birth time $t_\mathrm{b}$          & 2.047 & 2.042 & 2.038 & 1.93  & 1.84    & 2.11  & Ma \\
Initial porosity $\Phi_0$          & 0.281 & 0.301 & 0.314 & 0.301 & 0.299   & 0.300 &    \\
Peak central temperature           & 1248  & 1282  & 1320  & 1402  & 1516    & 1222  & K  \\[.1cm] 
            &  &  & \multicolumn{2}{c}{\bf Meteorites} \\
LON 94100  & \\
\quad burial depth                 & 20.1  & 17.2  & 15.3  & 13.8  & 12.5    & 20.1  & km \\
\quad peak temperature             & 1112  & 1118  & 1131  & 1159  & 1199    & 1098  & K  \\
Neuschwanstein & \\
\quad burial depth                 & 20.5  & 17.8  & 15.7  & 14.6  & 13.0    & 20.6  & km \\
\quad peak temperature             & 1117  & 1129  & 1141  & 1180  & 1212    & 1103  & K  \\
Khairpur & \\
\quad burial depth                 & 20.1  & 17.2  & 15.3  & 13.5  & 12.0    & 20.3  & km \\
\quad peak temperature             & 1112  & 1117  & 1132  & 1151  & 1185    & 1100  & K  \\
Blithfield & \\
\quad burial depth                 & 20.2  & 17.3  & 15.4  & 14.3  & 12.7    & 20.3  & km \\
\quad peak temperature             & 1113  & 1119  & 1134  & 1174  & 1205    & 1100  & K  \\
Daniel's Kuil & \\
\quad burial depth                 & 20.0  & 17.1  & 15.1  & 13.6  & 12.1    & 20.0  & km \\
\quad peak temperature             & 1112  & 1116  & 1128  & 1153  & 1190    & 1096  & K \\
\\
Quality function $\chi^2$          & 0.492 & 0.696 & 0.624 & 0.489 & 0.492   & 0.629 &   \\
\bottomrule
\end{tabular*}

\end{table*}
%--------------------------

The metamorphic temperatures corresponding to the different petrologic types are not well known for the enstatite chondrites \citep[see, for example, ][]{Hus06}. The latest study of this problem seems to be the work of \citet{Zha96}.

Their results for the metamorphic temperatures of EL6 chondrites are $800^\circ$C to $1\,000^\circ$C for the KQEOT thermometer.  However, there remained the uncertainty, that the KQEOT thermometer was set by pre-parent body events (e.g., chondrule formation, condensation). Other thermometers indicated lower temperatures, but were considered to possibly reflect later closure during slow cooling rather than maximum metamorphic temperatures. On the other hand, for EH6 chondrites, 
\citet{Zha96}  concluded $800^\circ$C--$1\,000^\circ$C as maximum metamorphic temperatures. A minimum value of $800^\circ$C for EL6 chondrites is also supported by the observation that all EL6 chondrites show well-defined I-Xe ages demonstrating that the closure temperature of $1\,000\pm100$ K was considerably exceeded during parent body metamorphism.  Hence, we have to require that the maximum of the central temperature of the parent body should exceed a lower limit of about $T_\mathrm{c,min}=1\,200$ K since the real EL6 chondrites probably originate not just from the central region of the body but somewhere else between the centre and the surface where temperatures are less. This requirement excludes models that correspond to minima in the $\chi^2$-surface found for $\chi_0=0.25$ and $\chi_0=0.30$ at formation times $t_\mathrm{b}\gtrsim2.15$ Ma. 

With respect to melting we have to consider three different major components in  enstatite chondrites: enstatite, plagioclase with about 15 mol\% anorthite, and the Fe,Ni-FeS complex. The melting of these three components as isolated substances starts with the onset of eutectic melting of FeS at a temperature of 1260 K \citep{Fei97}. At 1370 K the binary solid solution plagioclase starts melting which extends up to 1600 K, and at 1830 K pure enstatite melts. For the real mixture as found in enstatite chondrites the melting behaviour is more complex because a number of minor components are present and the composition of the components changes by equilibration between the various components. A laboratory investigation on melting of an EH chondrite is described in \citet{McC99} which shows that the basic features of the melting process for a chondrite are unchanged, except that the onset of melting of enstatite is lowered to a temperature of $\sim1\,750$ K because of changes in composition during heating of the mixture \citep{McC99}. Melting temperatures for Na-rich plagioclase in meteorites could also vary depending on the mineralogy of specific meteorite classes \citep[e.g., ][]{Col20}. Also, the melting temperature of Fe,Ni-FeS decreases with increasing Ni concentration. 

Figure~\ref{FigTc} shows as dotted lines for which radius-birthtime combination the maximum central temperature just equals the temperature required for the onset of melting of the main three components of the enstatite mineral mixture. The assumed temperatures are, however, to be considered rather as upper limits for the onset of melting. Since known EL6 chondrites do not show indications of enstatite melting we have to require that the maximum of the central temperature during the thermal evolution of the body does not exceed a limit temperature of $T_\mathrm{c,max}\sim1\,750$ K. This requirement excludes models with birthtimes significantly earlier than 1.65 Ma. An inspection of Fig.~\ref{FigThermEvo} then shows us that models with a high initial porosity $\Phi_0=0.40$ and $\Phi_0=0.35$ take a minimum of $\chi^2$ for models with a too early birthtime of the body so as to be compatible with the restriction for the maximum central temperature. Only the models with initial porosities around the value of 0.30 seem to be compatible with that requirement. Enstatite chondrites also seem to show no indications for extensive melting of plagioclase. 

%------------------------------------------------------------------------------
\subsection{Best-fit model}

The optimum fit of the available thermochronological data is determined by minimisation of $\chi^2$ with respect to the fundamental model para\-metres $R$ and $t_\mathrm{b}$. Besides these two principally unknown parametres there are additional parameters that determine the properties and evolution of the parent body but which are presently not known or for which one can give only crude estimates, though they are in principle determined by the physics of the planetesimal formation process and the properties of the accretion disk at the location and birthtime of the body. This holds for instance for the initial and surface temperatures, $T_\mathrm{i}$ and $T_\mathrm{s}$, respectively (cf. Sect.~\ref{SectIniSrfTem}), or the initial porosity of the granular material, $\Phi_0$. If sufficient data material is available, as is the case for the H chondrites, it can be attempted to fix them by including their value into the model optimisation process. For the EL chondrites our data-basis is very small and in principle an attempt to constrain further parameters by model-fitting should be avoided. Nevertheless, we check, which value of $\Phi_0$ is required for a good fit of the data by treating $\Phi_0$ as a free parameter that is determined by the optimisation process.
 
%-----------------------------------
\begin{figure*}
\centering

\includegraphics[width=0.8\textwidth]{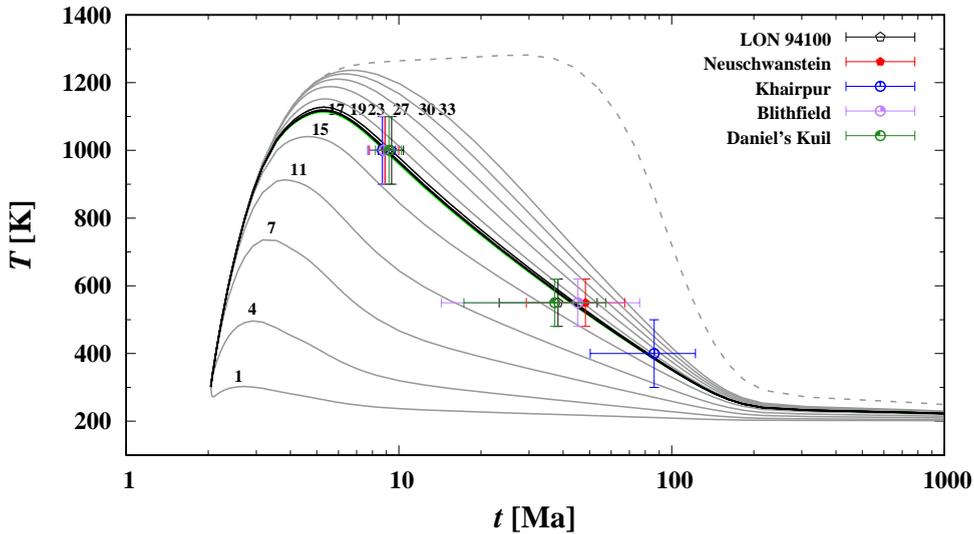}

\caption{
Temperature evolution at selected depths below surface for the optimized thermal evolution model (solid grey lines), and at the centre of the body (dashed grey line) for model (B) from Table \ref{TabOptMod}. The temperature evolution at the burial depth's of the meteorites are shown as solid black lines; they nearly coincide. The data-points corresponding to the closure-time and closure-temperature for the different meteorites and thermochronological systems are shown as symbols with error bars. The time corresponds to time elapsed after CAI formation. 
}

\label{FigThermEvoFit}
\end{figure*}
%-----------------------------------

In Fig.~\ref{FigThermEvo} (upper right panel) there are several local minima for birthtimes $<2.15$ Ma. For each of these, we give the model parameters and results in  Table~\ref{TabOptMod}, assuming an initial temperature of 300 K. For our preferred model (B, plotted in Fig.~\ref{FigThermEvoFit}) we also varied the initial temperatures (250 and 350 K) to show the influence of this parameter. Although the initial temperature could have possibly been higher, we demonstrate in the appendix (section \ref{SecTempDepMod}) that higher initial temperature cause only slight shifts to later formation times, while other results remain virtually unaffected. For the initial porosity, values between 0.25 and 0.4 were allowed and the allowed range for the radius $R$ and birthtime $t_\mathrm{b}$ is as in Sect.~\ref{SectChi}. The optimisation was done by a genetic algorithm.

It can be verified from Table~\ref{TabOptMod}, that each of the local minima represents a viable solution for certain parameters of birthtime, radius and porosity.  Table~\ref{TabOptMod} shows the results for initial radius $R_\mathrm{pc}$ before compaction, final radius $R$ after compaction, birthtime $t_\mathrm{b}$, and initial porosity $\Phi_0$ for the best-fit models and the corresponding values, $\chi^2$, of their quality function. Also given are the peak central temperature of the body, the burial depths of the meteorites, and the peak temperature reached at that location.  The main period of size reduction from pre-com\-pac\-tion radius $R_\mathrm{pc}$ fo final radius $R$ occurs around $1.5$ Ma after formation of the body. 

For all the six models the fit quality $\chi^2$ is acceptable. This means that it is not possible to decide on the basis of the fit quality alone which one of the different possible models reproduces the experimental data on the meteorites at best and possibly represents the properties of their parent body. 

Notably, the initial porosity for all models turns out to be very close to $\Phi_0=0.30$, a value which would be obtained for a random closest packing of equal sized chondrules with a small admixture of much smaller matrix particles. This result closely matches with the observation that primitive EL3 chondrites are mainly composed of chondrules and are matrix-poor.  The variation of the best-fit results for $T_\mathrm{c}$, $R$, and $t_\mathrm{b}$ with the assumption on $T_\mathrm{i}$ is small such that these can be considered as almost independent of $T_\mathrm{i}$ (see also appendix \ref{SecTempDepMod}).

The difference between the models of the local minima is mainly birthtime, ranging from 1.84 to 2.11 Ma after CAIs. A later birthtime correlates with lower central and EL6 maximum temperatures, with a smaller parent body radius and and with a deeper burial depth of the EL6 chondrites. While metamorphic temperature in all models (1\,100--1\,200 K) can be reconciled with the -- quite rough -- estimates available for EL6 chondrites, the central core peak temperature (1\,220-1\,520 K) allows incipient Fe,Ni-FeS eutectic melting in 2-3 cases and incipient plagioclase melting in only 1-2 cases. For a body with composition corresponding to chondrite types H, L, and LL a central temperature of 1\,500 K would be above the li\-qui\-dus of the silicate minerals in ordinary chondrites (ca. 1\,430 K) and the corresponding parent bodies would likely be differentiated. This is not the case, however, for a body of the characteristics of the parent body model (E) because of the significantly higher melting temperature of the nearly-pure enstatite ($\gtrsim1\,700$ K). 

%-----------------------------------
\begin{figure}
\centering

\includegraphics[width=0.45\textwidth]{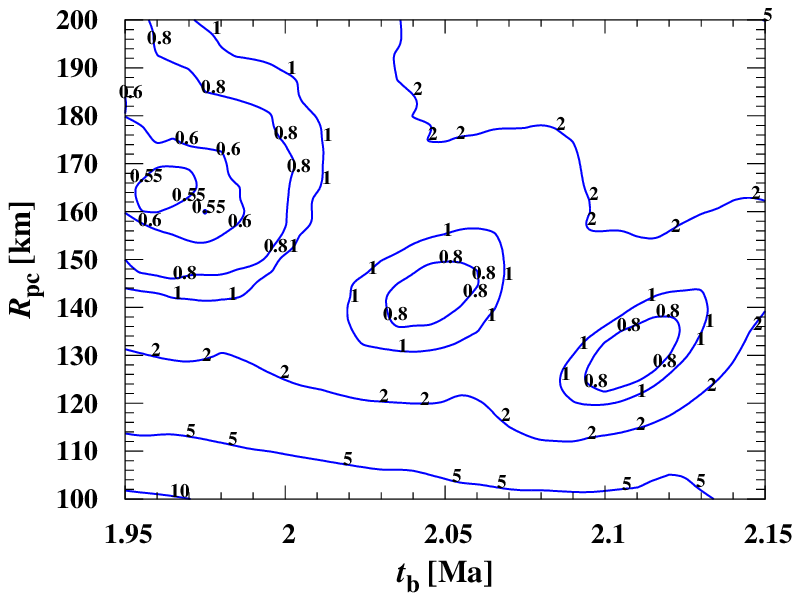}

\includegraphics[width=0.45\textwidth]{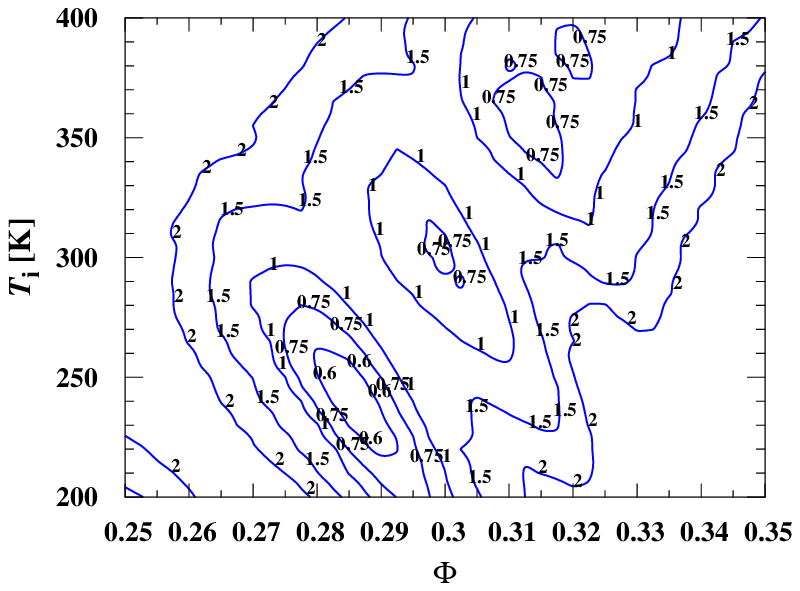}

\caption{
Variation of $\chi^2$ with initial radius $R_\mathrm{pc}$ and formation time $t_\mathrm{b}$ of the body (top panel) if an initial temperature of $T_\mathrm{i}=300$ K and initial porosity of $\Phi_0=0.30$ are assumed, and variation of $\chi^2$ with prosity $\Phi_0$ and initial temperature $T_i$ (bottom panel) in case that radius and formation time are fixed to the value corresponding to model (B) in Table \ref{TabOptMod}.
}

\label{FigVarChi}
\end{figure}
%-----------------------------------

It seems not possible to decide on the basis of the properties of the models which  models could represent the properties of the parent body of the EL chondrites. The only clear difference would be if EL chondrites from deeper parts of the parent body would be available where maximum temperatures at their burial depths exceed the liquidus temperature of plagioclase for the case that the models with higher central temperature apply. In that case the presence or absence of indications for plagioclase melting would allow a decision between the models. 
   
In fact, a few enstatite meteorites are described in the literature \citep{Boe14, Uri16a,Uri16b} which according to their composition seem to be related to EL chondrites and show indications of Fe,Ni-FeS melting and incipient plagioclase melting, but without indications for significant loss of melts. It is not quite clear whether the observed melting is due to high temperatures at their primordial location during the thermal evolution of the parent body or to impact heating of surface material. In case of Zak{\l}odzie a Al-Mg age of only 5.4 Ma after CAI formation was derived by \citet{Sug08} and this was ascribed to early $^{26}$Al induced thermal metamorphism, while other studies advocated early impact heating for this particular meteorite \citep[e.g. ][]{Prz05}. In \citet{Boe14} it is noted that the anomalous E chondrites studied by them compositionally are related to the EL chondrites and texturally resemble primitive achondrites indicating thermal and not impact heating. 

We take model (B) from Table~\ref{TabOptMod} to show some properties of the best-fit models. Figure \ref{FigThermEvoFit} demonstrates how the model fits the meteoritic data. The temperature variation with time of model (B) at the burial depths of the five meteorites are shown as black lines, the temperature variation for a number of other depth's as solid grey lines, and the central temperature as dashed grey line. The closure times and the closure temperatures of the corresponding thermochronometers are shown for the five meteorites  by different symbols and the corresponding error-bars. The $T(t)$ curves at the burial depths of the meteorites closely match the meteoritic data points well within their error bars. This shows that it is possible to consistently interpret the meteoritic data within the frame of the onion-shell model. 

The upper panel of Fig.~\ref{FigVarChi} shows in more detail how the quality function $\chi^2$ varies (at given values $\Phi_0=0.3$ and $T_\mathrm{i}=300$ K) with radius $R$ and birthtime $t_\mathrm{b}$ of the body around the optimum values corresponding to model (B). This shows that for each local minimum there is only a rather narrow time-window of about 0.02 Ma for the possible value of the birthtime of the parent body in order that the resulting thermal evolution model is consistent with the observed cooling history of our set of EL chondrites. However, the possible variation over all local minima is larger, c. $\pm0.08$ Ma, corresponding to precisions of some short-lived nuclide chronometries. The radius of the parent body is much less restricted. A variation by up to 20 km seems to be possible within local minima, and $\pm20$ km for all models (Table~\ref{TabOptMod}) involving only minor partial melting.

The lower panel of Fig.~\ref{FigVarChi} shows the variation of the quality function $\chi^2$ with assumed initial temperature $T_\mathrm{i}$ and porosity $\Phi_0$ for models with a radius and birthtime corresponding to model (B). This demonstrates in more detail what we noticed already above: The models only weakly depend on the precise value of the initial temperature, but the fit quality depends significantly on the initial porosities and fixes this to a value of appproximately $\Phi_0=0.3\pm0.05$, a value, that is also expected on physical grounds.  

%-----------------------------------
\begin{figure}
\centering
\includegraphics[width=.49\textwidth]{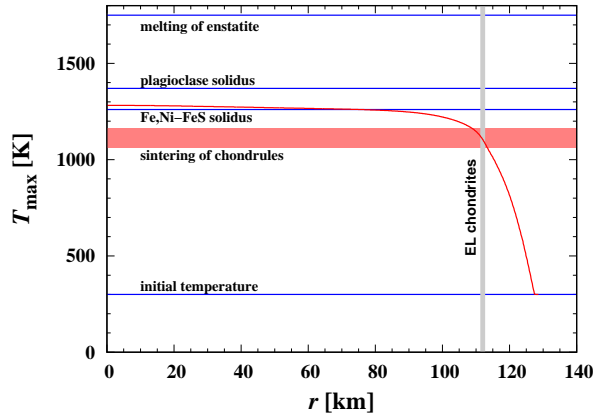}

\caption{
Radial variation of the maximum temperature achived during the evolution from birthtime up to 3 Gyr  within the parent body (red line) for model (B). Also indicated are the onset of melting of Fe,Ni-FeS, plagioclase, and of enstatite (the tempatures are only rough estimates, see text).  The wide red bar indicates the temperature range over which the enstatite chondrules are compacted (porosity reduction from 0.3 to 0.05). The vertical grey bar indicates the typical burial depths of the EL6 chondrites.  
}

\label{FigMaxT}
\end{figure}
%-----------------------------------

Figure~\ref{FigMaxT} shows for the case of model (B) the peak temperature to which the material was heated at radius $r$ during the first three Gyr of thermal evolution of the body. For comparison the temperatures are indicated for onset of melting of the major components of the enstatite chondrite material (enstatite, plagioclase, Fe-Ni and FeS) and the temperature range over which compaction of the granular material from the initial value $\Phi_0=0.3$ to a value of $\Phi=0.05$ (closure of pore space) occurred.

Because of the relatively high activation energy for deformation creep in enstatite the compaction occurs at significantly higher temperatures as for H and L chondrites (1\,150 K vs. 1\,000 K). In the model our EL6 meteorites are located within the layer where the loss of void space and the transition to a strongly solidificated material occurs. In this respect our model result for compaction is consistent with the compacted structure typical for  petrologic type six.

According to the model, the peak temperature suffered by the material of the EL6 chondrites during thermal evolution does not exceed the temperature of eutectic melting of the Ni,Fe-FeS sytem as estimated for the average composition of the nickel-iron alloy in EL chondrites, or the melting temperature of plagioclase. Under these circum\-stan\-ces no melting would be observed. This is consistent with a lack of observations of noticeable melting in the EL6 chondrites. 

Incipient melting, also of plagioclase, as is characteristic for primitive achondrites would be observed in models C, D, and E if meteorites originating from deeper burial depths (e.g. $>25$ km in our model E) on the parent body would be found \citep[which possibly already happened ][]{Boe14}. No differentiation into a metallic iron core and a silicate mantle would occur for the parent body within the frame of our model. The central temperature stays below the melting temperature of silicate such that no large-scale separation of silicate and iron melt is possible.

%********************************************************************

%********************************************************************

\section{Discussion}
\label{SectRes}

The comparison of our model with thermochronological data shows that solutions can be found that satisfactorily describe the thermal evolution of the EL chondrite parent body. In the following, we will discuss some implications and the validity of of our model asumptions and results, e.g., the validity of the asumption of instantaneous accretion, the EL parent body accretion age, its size, possible melting in the interior, and some implications of the nearly identical thermal history of various EL6 chondrites considered in this study.

\subsection{Validity of the asumption of instantaneous accretion}

Although it is clear that the initial stages of planetesimal growth are fast (see section~\ref{SectIniSrfTem}) when compared to decay time scales of $^{26}$Al, the most important heating nuclide, final stages of accretion may have been delayed. As the layering depths of the EL6 chondrites are relatively shallow, our model of thermal evolution assuming instantaneous accretion requires some discussion. In our models (Table~\ref{TabOptMod} - without model (E) allowing incipient plagioclase melting) the burial depths of EL6 chondrites are 15--20 km. It is important to note that maximum metamorphic temperatures (producing type 6 material) are mainly governed by accretion time, while the style of cooling (constrained by geochronological data) is determined mainly by layering depth. To be more exactly, ``layering depth'' means ``final layering depth after nearly complete release of decay heat''. This means, the layering depths in our model can be varied in a way that growth of the upper kilometre sized surface layers may have lasted up to 2 Myr longer, because $^{26}$Al decay heat needs 2--3 half-lifes to be released completely in the major (interior) mass of the body. Hence, the precursor material may have been layered more shallow in the beginning, however, reaching the peak metamorphic temperature of type 6 and cooling according to chronological data requires 15--20 km layering depths only c. 5 Ma after CAIs. This can be deduced from our Fig.~\ref{FigThermEvoFit}, which shows that the onset of cooling does not start before c. 5 Ma after CAIs for type 6 material. This time is, however, close to the disk dissipation age \citep{Hai01, Pfa15}, so significant pebble fluxes to feed asteroidal-sized bodies seem unlikely to occur \citep{Kla20}. Another line of evidence is deduced from various models for the H chondrite parent body. In this case, there are exceptionally high-quality chronological data constraining its thermal evolution, that also point at nearly instantaneous accretion within $< 1$ Ma \citep{Hen13,Mon13}. Particularly, these data comprise near surface sited low petrologic types (type 4) which are derived from burial depths as low as 5--7 km \citep{Hen13}, and contain metamorphic minerals with $^{26}$Mg excess from in situ $^{26}$Al decay. Together with the Pb-Pb, Ar--Ar and $^{244}$Pu fission track cooling ages \citep{Tri03}, this requires both fast accretion and fast cooling of the outer shell (within a few Ma) and is not in favor of a protracted accretion scenario.

\subsection{Accretion age of the EL chondrite parent body}

The range of accretion age of our EL chondrite parent body of 1.8--2.1 Ma after CAIs is comparable with other - H and L - chondrite parent bodies \citep{Hen13,Gai19}. In the case of the L chondrites \citep{Gai19}, individual chondrule ages were used to set an upper age limit on parent body accretion as chondrules must have formed via flash heating events before incorporation into their respective parent bodies. However, this requires high precision age dating of excellently preserved chondrules which have not been thermally disturbed during parent body heating. Such chondrules can only be found in the lowest petrologic types as, e.g., reported by \citet{Pap19}. In the case of EL chondrites, low petrologic types are rare, and so are chondrule age data.  \citet{Zhu20} found an age of $4565.7\pm 0.7$ Ma for enstatite chondrite chondrule formation, corresponding to  $1.6\pm0.7$ Myr after CAIs. This can easily be reconciled with a parent body formation age of 1.8--2.1 Ma after CAIs. 

\subsection{EL chondrite parent body size}

Our best fit models without plagioclase melting yield (final) parent body radii of 120--160 km, comparable to the H or L chondrite parent bodies. Previous estimates of the size of enstatite chondrite parent bodies were based on the total cooling time interval defined by type 6 chondrites. \citet{Bog10} argued that H6 ordinary chondrites record younger Ar-Ar cooling ages and accordingly longer cooling than enstatite type 6 chondrites, implying a possibly smaller E chondrite parent body. However, detailed parent body modelling demonstrates that asteroids cool shell by shell from the exterior, i.e., some shallow layered material may reach metamorphic temperatures nearly as high as the centre, but cools much more rapidly. In the case of ordinary chondrites, modelling yields layering depths for type H6 chondrites between 18 and 40 km in a $R=160$--200 km body \citep[e.g., ][]{Hen13,Hen16}, meaning that some H6 chondrites record cooling at higher burial depths, but not necessarily in a larger parent body. In contrast, the EL6 chondrites modelled here record the younger Ar-Ar ages due to shallow layering in c. 15--20 km depth in a $R=120$--160 km body. As noted already by \citet{Hop14}, the old EL6 Ar-Ar ages are either due ``to a smaller parent body of EL chondrites relative to the H chondrite parent body, or alternatively, to a shallower sampling depth of the meteorites''. Our modeling suggests that likely the latter is the case, although a smaller body size is likely for models that do not allow incipient plagioclase melting. Relatively shallow layering of EL6 chondrites also implies that the EL3, 4, 5 shells are confined to quite thin surface layers. This in turn could also partly explain the scarcity of these petrologic types.

As a final cautionary remark, we think that constraints on parent body size provided from samples by a range of layering depths --- as in the H and L chondrite asteroids --- represent stronger constraints than for the EL case. 

\subsection{Partial melting in the interior}

Among our models listed in Table 3, only few would allow incipient melting of Fe,Ni-FeS or plagioclase in the central parent body region. Indeed, there is hardly evidence for partial melting, not to mention melt migration in the case of EL chondrites. In the EL meteorite collections we know petrologic types ranging from 3 to 6, with a dominance of type 6. There are no achondritic meteorites of EL characteristics. While differentiated enstatite chondrites exist (e.g. aubrites or Shallowater) a  common origin from the EL parent body was excluded \citep[e.g., ][]{Kei89}. Only few enstatite meteorites with incipient melting (but no melt loss) are possibly related to EL chondrites \citep{Boe14, Uri16a,Uri16b}, e.g., Zak{\l}odzie. However, it is not clear, if its genesis is associated with strong metamorphic heating or impact metamorphism. Its old Al-Mg age of 5.4 Ma after CAIs, seems incompatible with a time temperature path experiencing $> 1\,400$ K and cooling to the Al-Mg closure soon afterwards.  Hence, if Zak{\l}odzie at all was related to the EL parent body, only impact excavation could explain its high heating degree and rapid cooling \citep[e.g., ][]{Prz05}.

\subsection{Collisional evolution}

In general, thermochronological data are more limited than in other favorable cases, e.g. the H chondrite asteroid \citep{Hen13,Hen16} or the Acapulcoite-Lodranite (A-L) parent body \citep{Neu18}, where up to 4 chronometers are available for a significant number of individual meteorites. Another factor limiting the modeling approach is the similarity of chronological data among the EL chondrite samples, as only for EL chondrites of petrologic type 6 reasonable data are available, which obviously stem from a limited depth range in their parent body. Layering depths vary only by $<1$ km for the five modelled EL6 chondrites (Table~\ref{TabOptMod}), and stem from somewhat shallow layers, consistent with estimates by \citet{Zha96b}.

A similar situation was found for the Acapulcoite- Lodranite parent body, for which modeled layering depths of acapulcoites were essentially indistinguishable, while lodranites obviously stem from only c. 3 km deeper shells \citep{Neu18}. This contrasts with various H chondrites of different petrologic types which were modelled to stem 
from depths varying by many tens of km \citep{Hen13,Hen16}. The question is why in some cases (EL and A-L meteorites) sampling seems to be restricted to a limited range of layering depths.

A possible reason is linked to the mechanism how small meteorites are sampled from initially large parental asteroids. Meteoroids have to be excavated from their parent bodies by collisions, and Monte Carlo simulations \citep{Tur79} indicate that typically several collisions are involved from fragmenting initially large asteroids to fragments of 
(observable) asteroid families, and finally to meter-sized meteoroids. Cosmic ray exposure ages demonstrate that the excavation of typical meter-sized meteoroids (later impacting on Earth) occurred no longer than a few tens of millions of years ago for stony meteorites. Hence, it is not astonishing that meteorite fluxes maybe dominated by a small restricted region of their initially larger parent bodies. From these comparatively small objects, several or maybe only few impacts may excavate meteorites which dominate the flux of this specific class to Earth.

Such a scenario is supported by the clustering of cosmic ray exposure (CRE) ages of acapulcoites and lodranites. CRE ages are indistinguishable at $5\pm1$ Ma \citep{Eug05}, indicating that possibly only one impact excavated most acapulcoites and lodranites 
from a c.~3~km sized fragment of the initially much larger A-L asteroid \citep{Neu18}. For the H chondrites CRE ages range up to 80 Ma indicating that continuous impact events excavated meteoroids from several fragments or rubble pile asteroids consisting of different fragments representing a wide layering depth of the initial H chondrite 
asteroid.

In the case of EL6 chondrites, an origin from a small, $<1$ km sized Apollo asteroid -- or an impact region of such size on a larger asteroid -- seems feasible. The CRE age distribution of the EL6 chondrites studied here (Blithfield: $30\pm5$ Ma; Khairpur: $37\pm5$ Ma; Daniel's Kuil: $33\pm5$ Ma \citep[][and references therein]{Cra81} ; Neuschwanstein: 47 Ma \citep{Zip10}; LON94100: $31\pm2$ \citep{Pat01} has four members compatible with a 33 Ma age, quite similar to the majority of EL6 CRE ages that cluster between 20 and 35 Ma  \citep{Pat01}). Although not as compelling as in the case of A-L meteorites, this may indicate excavation from a relatively small parent body fragment by a single low grade impact event. The occurrence of other petrologic types in the fall statistics of enstatite chondrites maybe explained by stemming from other small fragments of the EL parent asteroid.

It should be noted here that the eventual impacts excavating meter-sized meteoroids from parent body fragments are usually low energetic impacts that are not capable of resetting radioisotope chronometers. This can be deduced from many observations that CRE ages (defined by nuclides accumulating by cosmic ray interactions) are always much younger than radioisotopic ages set by thermal events exceeding respective closure temperatures. While most CRE ages of stony meteorites do not exceed 100 Ma, most radioisotopic ages are related to parent body metamorphism in the early solar system c. 4.5 Ga ago  \citep[e.g., ][]{Tri09}. Occasionally, highly energetic impact events can be recorded, however, mostly by chronometers with low closure temperatures. For example, many U/Th--He and Ar-Ar ages indicate a major catastrophic collision on the L chondrite parent body 470 Ma ago \citep{Kor07}. However, such events set by asteroid collisions are usually accompanied by significant shock metamorphism, leaving typical signatures in meteorites like conversion of plagioclase to maskelynite, production of localised melt veins and pockets, and transformation of minerals to high-pressure polymorphs \citep{Sto91}. In studies aiming at the chronology of early parent body cooling, shock-metamorphosed samples are avoided, usually only samples with low shock stages 1 or 2 are considered \citep[e.g., ][]{Tri03}. While the effects of impact metamorphism can be safely identified in cases of localised melt veins and pockets, problems may occur in case of ascribing more global melting phenomena to either internal parent body heating or to external impact heating by large impacts, e.g. such as in the case of the primitive enstatite achondrite Zak{\l}odzie (see above). 

\subsection{Implications for the EH chondrite parent body}

A similar modelling study for the EH chondrite parent body is much more complicated or even impossible due to the scarcity of chronological data resulting from parent body cooling. For example, low temperature chronometers frequently show secondary disturbances or complete reset by impact metamorphism, e.g., a $< 1$ Gyr Ar-Ar age for EH4 Bethune \citep{Bog10}, or Rb-Sr ages for EH3 Qingzhen of $2.12\pm 0.23$ Gyr and for EH3 Yamato 06901 of $2.05\pm0.33$ Gyr \citep{Tor93}. 

High temperature chronometer are more likely to record the early parent body cooling history. In the case of the I-Xe chronometer, we indeed have a remarkable coherence of ages of four EH Chondrites (EH4 Abee and Indarch, EH4/5 EET96135 and EH5 St. Marks) which are all ca. 4 Ma older than the EL6 chondrites \citep{Hop16,Ken88,Bus08} indicating I-Xe closure only 6 Ma after CAIs. Other constraints can be obtained from the fact that for EH chondrites there is a preponderance of low petrologic types but nevertheless few type 6 are known. A maximum metamorphic temperature of type 6 constrains the 
formation time to similar values as the EL chondrite parent body, i.e. 2.0-2.1 Myr after CAIs. Such a formation time would be compatible with the EH chondrule age of $1.6\pm0.7$ Myr after CAIs by \citet{Zhu20}. I-Xe closure (at 1000 K) c. 4 Myr earlier than EL6 types could be achieved in a layering depth of 15 km (compare Fig. \ref{FigThermEvoFit}), implying a maximum metamorphic temperature c. 100 K lower than type 6. 

In contrast the size of the parent body can only be deduced from the cooling curve shape, e.g., a steep cooling curve implies lower radii, a shallower cooling curve larger radii. Hence, the EH parent body size can only be constrained by an undisturbed cooling age of a low temperature chronometer like Ar-Ar, which seems presently unavailable and must await future studies of samples undisturbed by secondary impact metamorphism.

%****************************************************************************
\section{Concluding remarks}

We attempted to reconstruct the parent body of the group of enstatite chondrites with low iron content (type EL). This is done by comparing information on the cooling history of individual EL meteorites provided by Ar-Ar and I-Xe ages of meteorites with model calculations for the internal constitution and thermal evolution of big planetesimals with material properties characteristic for enstatite chondrites. 

We use for this comparison chronological data previously determined by  us \citep{Hop14,Hop16} and other studies \citep{ Ken81, Ken88, Bus08, Bog10} and selected meteorites for which at least two reliable such age determinations exist for thermochronological systems with different closure temperatures. We additionally determined the frequency factor and activation energy for Xe-diffusion in the enstatite host material to determine the closure temperature of the I-Xe system. Because of the relative rarity of E chondrites there are only five meteorites (LON 94100, Neuschwanstein, Khairpur, Blithfield, Daniel's Kuil) with data useful for our purposes, all of petrologic type six. This is far less than for the much better studied H chondrites and unfortunately covers only a limited range of peak metamorphic temperatures. This number is just sufficient to pin down the two most important parameters of the parent body, its radius and its formation time, by fitting models for the internal constitution and evolution of small bodies from the asteroid belt to the meteoritic record. 

The model calculations follow closely the methods applied in our previous work on H chondrites \citep{Hen12,Hen13,Gai15,Gai18}, L chondrites \citep{Gai19}, and Acapulcoites and Lodranites \citep{Neu18}. 
Based on the assumption of a spherically symmetric body that is heated by decay of long and short lived radioactives, in particular by $^{26\!}$Al, and formed on a timescale short compared to the initial heating period, we solved the heat conduction equation coupled with equations for the compaction of the initially chondrule-dominated material, considering the temperature and porosity dependence of heat conductivity and the temperature dependence of heat capacity. 

By varying the radius, the formation time (after CAI formation), and the initial porosity of the material we determined the parameter combination which fitted the  thermo\-chro\-nological data as well as possible. Probably because of the restricted number of meteorites to compare with, more than one solution is possible. Two models are consistent with no incipient melting on the EL chondrite parent body, with accretion ages of 2.11 and 2.04 Ma after CAIs, burial depths of  17 and 20 km, and peak temperatures of 1\,100 K and 1\,120 K. Two more models (1.93 and 1.84 Ma after CAIs) allow incipient to minor plagioclase melting (but no melt migration) and could apply if recently identified enstatite chondrites with properties resembling primitive achondrites were derived from the EL chondrite parent body. The parent body radius imposed by our models is less well constrained varying between 120 and 210 km, however, the initial porosity is close to 30\% in all cases. The  initial temperature is only weakly restricted by the model fits, but a range between about 250 K and 350 K is suggested.

\medskip
{\small{\bf Acknowledgement:}
We thank two unknown referees for their constructive comments which helped to improve the manu\-script. This work was performed as part of a project of `Schwerpunktprogramm 1833', supported by the `Deutsche For\-schungs\-gemeinschaft (DFG)'.  MT and JH acknowledge support by the Klaus Tschira Stiftung gGmbH. This research has made use of NASA's Astrophysics Data System.
}

%********************************************************************

\appendix

\section{Material properties}

\label{SectMaterial}

For modelling the thermal evolution of an asteroid we require information on a number of properties of its material: the heat capacity, $c_V$, heat conductivity, $K$, heating rate, $h$, and deformation properties during sintering. In the following we describe how these data are determined.

%-------------------------------------------------------------------------------
\subsection{Composition}

The properties of the material are determined by its chemical and mineralogical composition which can be derived from the analyses of the composition of the enstatite meteorites. The enstatite chondrites consist of a granular material composed of a complex mixture of about 0.1 to 1 mm sized roundish chondrules composed themselves of a mixture of enstatite and some less-abundant Al-Ca-Na-silicate minerals, and a small fraction of matrix material consisting of $\mu$m-sized dust grains (if ever) and some fraction of metallic iron and iron-sulphide grains filling the voids between chondrules.

%---------------------------
\begin{table}
\caption{
Dominating mineral components in enstatite chondrites, their bulk densities $\rho$, mass-fractions $X$, and volume fractions $f$.
}
\label{TabComp}

\begin{tabular*}{\tblwidth}{@{}LLLLL@{}}
\toprule
species     & composition & $\varrho$ & $X$ & $f$ \\
\midrule
            & \multicolumn{3}{c}{EH} \\
Olivine     & ---                         &      &      &       \\
Pyroxene    & En$_{100}$                  & 3.20 & 0.56 & 0.666 \\
Plagioclase & Ab$_{81}$Or$_{4}$An$_{15}$  & 2.64 & 0.10 & 0.144 \\
Nickel-iron & Fe$_{92}$Ni$_8$             & 7.90 & 0.25 & 0.12  \\
Troilite    & FeS                         & 4.91 & 0.09 & 0.07  \\[.1cm]
\multicolumn{2}{l}{Bulk density}          & 3.78 & \\
            & \multicolumn{3}{c}{EL} \\
Olivine     & ---                         &      &      &       \\
Pyroxene    & En$_{100}$                  & 3.20 & 0.64 & 0.726 \\
Plagioclase & Ab$_{81}$Or$_{4}$An$_{15}$  & 2.64 & 0.10 & 0.137 \\
Nickel-iron & Fe$_{94}$Ni$_6$             & 7.90 & 0.20 & 0.092  \\
Troilite    & FeS                         & 4.91 & 0.06 & 0.044  \\[.1cm]
\multicolumn{2}{l}{Bulk density}          & 3.63 & \\
\bottomrule
\end{tabular*}

\end{table}
%--------------------------

Average values for the mole fractions of the components in EH meteorites are estimated from data given in \citet{Zip10,Wei12} and \citet{vSc69}. Our model data are listed in Table~\ref{TabComp}. These data are based on data referring to EL meteorites of low and high petrologic types and therefore represent some kind of average composition with respect to the main components for all petrologic types. In fact, there are compositional differences between primitive and highly equilibrated meteorites with respect to minor components in the mixture, but this would cause only minor changes in the material properties if considered. The bulk density
\begin{equation}
\varrho_\mathrm{b}=\sum_i X_i\rho_i 
\end{equation}
calculated for this composition is $\varrho_\mathrm{b}=3.63$ g\,cm$^{-3}$, while for comparison \citet{Fri15} gives measured values for $\varrho$ of $3.48\pm0.18$ for Hvittis and $3.62\pm0.01$ for Pillistfer.

%-------------------------------------------------------------------------------
\subsection{Heat capacity}

The heat capacity and conductivity could in principle be measured directly from meteoritic specimens, but in case of EL chondrites such data seem not to be available. We therefore have to estimate the relevant data from laboratory measured data for the constituents of the chondritic material. The heat capacity per unit mass for the mixture of minerals and iron encountered in enstatite chondrites is calculated as
\begin{equation}
c_V=\sum_i X_i c_{V,i}\,.
\end{equation}
from laboratory measured heat capacities $c_{p,i}$ per unit mass of the mixture components (we neglect the small difference between $c_p$ and $c_V$ of solids) and their mass-fractions $X_i$ in the mixture. We consider only the most abundant components and an average composition of those components which are solid solutions as specified in Table~\ref{TabComp}. Details are described in \citet{Gai12}. 

%-----------------------------------
\begin{figure}
\centering
\includegraphics[width=.49\textwidth]{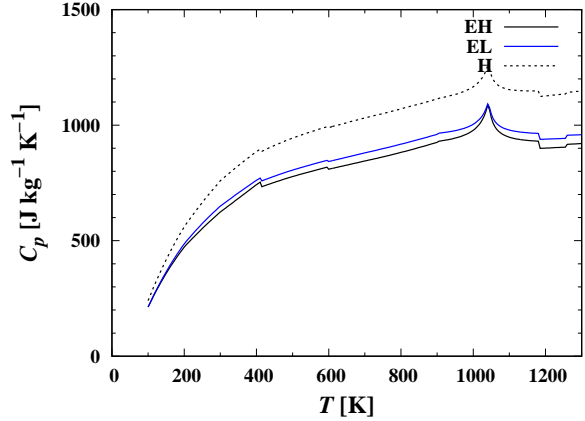}
\caption{
Temperature variation of the specific heat $c_V$ for the composite material in EH and EL chondrites. For comparison, also the specific heat for H chondrites is shown.
}

\label{FigCV}
\end{figure}
%-----------------------------------

Figure~\ref{FigCV} shows the resulting temperature variation of the calculated specific heat, $c_V$, for EL chondrites. No experimentally determined data seem to be available which can be compared to our model data to check the accuracy of the approach. The values cited in \citet{Soi20} and in \citep{Ope12} as taken from \citet{Yom83} seem, in fact, to be theoretical estimates because \citet{Yom83} does not contain data on enstatite chondrites. A laboratory measured value given in \citet{Rud81} seems be either erroneous because it is significantly higher than for the pure substances in the mixture, or their specimen had an untypical composition. Since the specific heats of H and L chondrites calculated by the same method compare well with laboratory data \citep{Gai19} we believe that our results for enstatite chondrites are also reliable. 

%-------------------------------------------------------------------------------
\subsection{Heat conductivity}

\label{SectHeatCond}

Measured data presently seem to be only available for the heat conductivity of the EL6 meteorite Phllistfer studied by \citet{Ope12}. This measurement is limited, however, to the low temperature regime $T<340$ K which does not cover the important high-temperature range up to at least 1\,400 K relevant for thermal models of meteorite parent bodies. Hence we have to determine the heat conductivity from a theoretical model for the mineral mixture of enstatite chondrites.

%-----------------------------------
\begin{figure}
\centering
\includegraphics[width=.49\textwidth]{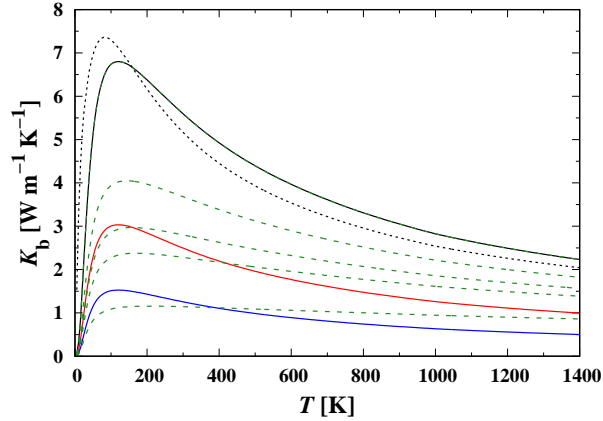}
\caption{
Temperature variation of the bulk heat conductivity $K$ for the composite material in EL chondrites (black solid line), and for porous material with porosities $\phi=0.25$ (red solid line) and $\phi=0.35$ (blue solid line). For comparison also the bulk heat conductivity of H chondrite material is shown (short dashed black line). 
}

\label{FigKth}
\end{figure}
%-----------------------------------

The heat conductivity depends on the mineralogical composition of the chondritic material and additionally it depends strongly on the porosity of the material because any void represents an obstacle for heat flow. It can be described by the heat conductivity of the bulk material, $K_\mathrm{b}$, and a correction factor, $f(\phi)$, accounting for the reduction of the conductivity due to the presence of pores as in Eq.~(\ref{DefKpor}). 

First we consider the bulk heat conductivity $K_\mathrm{b}$ and its temperature variation. It is calculated by the model described in  \citet{Gai18} for matter with the average composition given in Table~\ref{TabComp}. The result is shown in Fig.~\ref{FigKth}. 

A comparison of this with real enstatite chondrite material is hampered for two reasons. First, measured heat conductivity data are available only for the two enstatites Abee (EH6) and Pillistfer (EL6) in the temperature range $T<340$ K. Second, experimental data for the heat conductivity for the components of the chondritic material (with the exception of iron) could only be found for the temperature range $T\gtrsim300$ K. For this reason a comparison of our heat conductivity model with laboratory masurements is presently only possible for the temperature range around 300 K.

In \citet{Gai18} the temperature variation of the components were approximated for the whole temperature range $0< T <1500$ K by the model of \citet{Cal59} which allowed a good approximations of the heat conductivities of H, L, and LL chondrite material. For the two enstatite chondrites, however, the model derived for the temperature dependence of $K$ for the mineral enstatite in \citet{Gai18,Gai19} matches the heat conductivities of the two enstatite chondrites at low temperatures $T\lesssim100$ K not particular well \citep[cf. Fig. 6 in ][]{Gai18}. This likely results from a lack of data for very low temperatures which forced us in particular to use a rather crude approximation for the Debye temperature. Fortunately, the temperature range $T\lesssim200$ K is not relevant for thermal evolution modelling of the enstatite chondrite parent body. In the important temperature range $T\gtrsim300$ K on the other hand, our heat conductivity model for enstatite reproduces the available experimental data \citep[cf. Fig. 2 in ][]{Gai18} at $T\ge300$ K reasonably well. 

Second, we consider the correction factor $f(\phi)$ accounting for the presence of pores. For this we use in our model calculations the approximation \citep{Hen16,Gai12}
\begin{align}
f_1(\phi)&=\ max\left(\,1-2.216\,\phi,\ 0\,\right) \label{FacKlow}\\
f_2(\phi)&=\ \mathrm{e}^{1.2-\phi/0.167} \\
f(\phi)&=\ \left(f_1^4(\phi)+f_2^4(\phi)\right)^{1\over4}\,.
\end{align}
This joins smoothly an approximation $f_1(\phi)$ applicable for porosities $\phi\lesssim0.35$ with an approximation $f_2(\phi)$ for porosities $\phi\gtrsim0.45$.

We note that our model fairly reproduces the measured value at 300 K for Pillistfer. The experimental value for Philistfer at 300 K is $K=5.14$ W\,m$^{-1}$K$^{-1}$ \citep{Ope12}, while for our model we obtain $K_\mathrm{b}=5.59$ W\,m$^{-1}$K$^{-1}$. The effective heat conductivity is essentially determined by that of the enstatite component ($K_\mathrm{b}=5.07$ W\,m$^{-1}$K$^{-1}$) because of its high abundance and by the nickel-iron component ($K_\mathrm{b}=33.8$ W\,m$^{-1}$K$^{-1}$) because of the high heat conductivity of metals. If we consider that the measured porosity of Pilistfer is $\phi\sim0.03$ \citep{Mac10} we obtain with Eqs. (\ref{FacKlow}) and (\ref{DefKpor}) a value of $K=5.22$ W\,m$^{-1}$K$^{-1}$ in reasonable agreement with the experimental finding.

%---------------------------
\begin{table}
\caption{
Data for calculating the heating rate.
}
\label{TabHeat}

\begin{tabular*}{\tblwidth}{@{}LLLLL@{}}
\toprule
isotope   & $X$     & $f$ & $E$ & $\tau$ \\
          &         &                    & [eV]  & [a]             \\
\midrule
$^{26}$Al & 0.010   & $5.1\times10^{-5}$ & 3.188 & $1.0\times10^6$ \\
$^{40}$K  & 0.00070 & $1.5\times10^{-3}$ & 0.693 & $1.8\times10^9$ \\
$^{60}$Fe & 0.248   & $1.1\times10^{-8}$ & 2.894 & $3.8\times10^6$ \\
\bottomrule
\end{tabular*}

\end{table}
%--------------------------

%-------------------------------------------------------------------------------
\subsection{Energy production}

Heat production within the body is assumed to be domi\-nated by the decay of radioactive nuclei. The heating rate per unit mass is
\begin{equation}
h=\sum_i X_if_iE_i\tau_i\,\mathrm{e}^{-t/\tau_i}\,,
\end{equation}
where the sum is over all isotopes, labelled with index $i$, contributing significantly to the overall heat production. Here, $X_i$ denotes the mass fraction of the corresponding element, $f_i$ the fractional abundance of the radioactive isotope at the time of formation of the solar system, $E_i$ the energy released by the decay of that nucleus to its stable end-product, $\tau_i$ the average lifetime of the radioactive nucleus, and $t$ is the time elapsed since solar system formation. 

The most important nuclei for internal heating of the body are $^{26}$Al, $^{40}$K, and $^{60}$Fe. Values for $f_i$, $E_i$, and $\tau_i$ are given in Table 4 of \citet{Gai12}. The mass-fractions $X_i$ of the elements are set to the average mass fractions in EL chondrites as given in Table 3 of \citet{Hop09}; they agree reasonably with the values given for Neuschwan\-stein in \citet{Zip10}. For the abundance of $^{60}$Fe we use the value given in \citet{Tan12} which seems also to apply for EL chondrites. At this small level of abundance of $^{60}$Fe, its contribution to heating is unimportant.

It is assumed that the initial distribution of the heat sources within the body is homogeneous and that they are not later re-distributed by, for instance, differentiation. 

%----------------------------
\begin{figure}
\centering
\includegraphics[width=.49\textwidth]{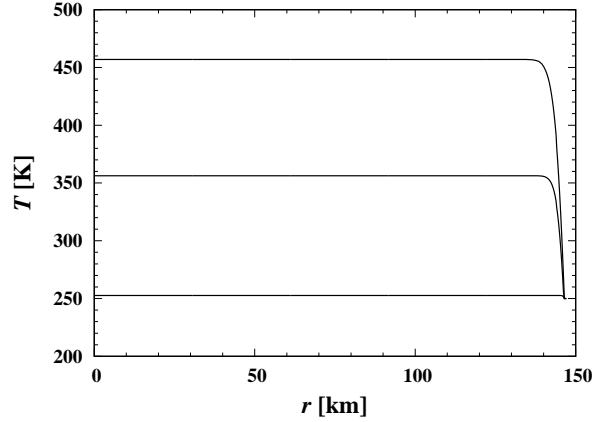}

\caption{Radial variation of temperature for model (A) from Table~\ref{TabOptMod} at 2\,000, 90\,000, and 200\,000 yrs after formation.}

\label{FigTemProf}
\end{figure}
%----------------------------

%-------------------------------------------------------------------------------
\subsection{Model dependence on initial temperature}
\label{SecTempDepMod}

During the initial first $10^5$ years during the evolution of asteroids with at least $\gtrsim100$ km radius, the radial temperature profile is rather flat. The  characteristic timescale  for temperature changes by heat conduction is $\tau=\kappa/d^2$ for a layer of thickness $d$ and heat diffusivity $\kappa=K/(\varrho c_p)$. This amounts to ca. $5\times 10^6$ yrs for $d=10$ km and a typical diffusivity $\kappa$ of $5\times10^{-7}$ (see Figs.~\ref{FigKth} and \ref{FigCV} and Table~\ref{TabComp}). During the initial heating phase, the spatial temperature-increase from the outer boundary temperature to the actual inner temperature caused by $^{26}$Al decay is limited to a layer of a few km thickness (see Fig~\ref{FigTemProf} for an example) . As a consequence, from two bodies with otherwise identical properties formed at instants $t_1<t_2$ with initial temperatures $T_1<T_2$, body 1 has reached just temperature $T_2$ at instant $t_2$ if the differences of the total heat contents of the body corresponding to the temperature difference equals the heat release by $^{26}$Al during the period from $t_1$ to $t_2$:
\begin{equation}
\int_V\int_{T_1}^{T_2}\varrho c_p(T)\mathrm{d}T\,\mathrm{d}V=\int_V\int_{t_1}^{t_2}\varrho h\,\mathrm{d}t\,\mathrm{d}V\,.
\end{equation}
(The integration over $V$ is over the volume of the body.) From that point on, body number one takes the same thermal evolution as body number two. Hence, different initial temperatures of otherwise identical bodies are equivalent to a shift in formation time, at least approximately, if they show the same properties at later times.

%-------------------------------------------------------------------------------
\subsection{Sintering}

\label{SectSint}

Sintering is modelled by the approach described in \citet[][cf. the references therein for the papers from which this model is adapted]{Gai15}. Only sintering by power-law creep is considered because matrix material has low abundance and creep dominates for the relatively big chondrules. 

The equations for sintering are more conveniently written in the variable
\begin{equation}
D=1-\phi
\end{equation}
which is the so called filling factor. This variable is used in what follows. The initial filling factor of the granular chondritic material is $D_0$. The compaction has to be treated somewhat differently for the initial stage where the pore-space is still interconnected, and the final stage where the pore network is closed. The transition between both cases occurs at a filling factor $D_\mathrm{cl}$.

The sintering is caused by an effective pressure acting between the contact areas of the chondrules. During the initial stage ($D_0\le D\le D_\mathrm{cl}$) this is given by
\begin{equation}
P_{1\mathrm{eff}}=P\,{1-D_0\over D^2(D-D_0)}+{3\gamma_\mathrm{sf}\over G}\,D^2{2D-D_0\over1-D_0}-p_\mathrm{g}
\end{equation}
and during the final stage ($D_\mathrm{cl}<D\le 1$) by
\begin{equation}
P_{2\mathrm{eff}}=P+{2\gamma_\mathrm{sf}\over G}\left(6D\over1-D\right)^{1\over3}-p_\mathrm{cl}\,{(1-D_\mathrm{cl})D\over(1-D)D_\mathrm{cl}}\,.
\end{equation} 
Here, $P$ is the pressure excerted on the granular material, $p_\mathrm{g}$ the gas pressure in the voids, $p_\mathrm{cl}$ the gas pressure in the pores at the moment of the closure of the pore network, and $G$ is the average radius of the granular units (the chondrules).

The variation of the filling factor with time is described during the initial stage by
\begin{equation}
{\mathrm{d}\,D\over\mathrm{d}\,t}=3.06\,\left(D^2D_0\right)^{1\over3}\left(D-D_0\over 1-D_0\right)^{1\over2}\,C(T)\left(P_{1\mathrm{eff}}\over3 P_0\right)^n
\end{equation}
and during the final stage by
\begin{equation}
{\mathrm{d}\,D\over\mathrm{d}\,t}=1.5\,{D(1-D)\over\left(1-(1-D)^{1\over n}\right)^n}\ C(T)\left(3P_{2\mathrm{eff}}\over2n P_0\right)^n\,.
\end{equation}
The quantity $C$ is 
\begin{equation}
C(T)=A\mathrm{e}^{-Q/(R_\mathrm{g}T)}\,,
\label{CreepDef}
\end{equation}
where $Q$ is the activation energy for creep, $A$ the frequency factor, $n$ is the power of the creep law, $R_\mathrm{g}$ is the gas constant, and $P_0$ is a reference pressure. These equations present a differential equation for the evolution of the filling factor during hot pressing that has to be solved with some initial condition.

%---------------------------
\begin{table}
\caption{
Data for modelling sintering of enstatite by power-law creep.
}
\label{TabSint}

\begin{tabular*}{\tblwidth}{@{}LLLLL@{}}
\toprule
Quantity         &         & Value & Unit \\
\midrule
Creep parameter  & $A$    & $1.1\times10^{8}$ & s$^{-1}$       \\
                 & $Q$    & $603\pm54$         & kJ\,mol$^{-1}$ \\
                 & $n$    & 3.0                &                \\
Reference pressure & $P_0$ & 1                 & bar            \\[.2cm]
Surface energy  & $\gamma_\mathrm{sf}$ & 1     & J\,m$^2$       \\
Chondrule radius & $G$    &  0.3               & mm             \\
Loosest random packing     & $D_0$  & 0.56               &  \\
Closure of void-space      & $D_\mathrm{cl}$ & 0.95      &  \\
\bottomrule
\end{tabular*}

\end{table}
%--------------------------

The specific material properties are described in this approach by $A$, $n$, and $Q$, which have to be determined from laboratory measurements. For enstatite chondrites sintering occurs under dry conditions. This is important because the activation energy for creep depends strongly on the oxygen fugacity and is the higher the dryer the conditions in the system are. Because of the long timescales for heating in asteroidal bodies sintering occurs in such bodies at rather low temperatures, typical $\sim 1\,000$ K, which fall into the stability field of orthoenstatite \citep[cf. Fig. 1 in ][]{Bys16}.

Creep deformation of (nearly) pure ortho-enstatite at different oxygen fugacities was studied in the laboratory. Over\-views on the available data on different materials of interest in particular for geophysical problems, but also required for modeling of asteroids, are given, e.g., by \citep{Bur08} and \cite{Kar13}. Two more recent studies dealing with ortho-enstatite are \citet{Bys16} and \citet{Zha20}, from which the second studied deformation under ``wet'' and ``dry'' conditions for which the hydrogen content of the samples were explicitly determined. The hydrogen content of enstatite chondrites according to \citet{Jar90} may be of the order of 0.1wt\%, which converts to a H/Si-ratio of the order of $10^{-4}$ which approximately correponds to the H/Si ratio of the experiments under ``wet'' conditions in  \citet{Zha20}. The coefficients for Eq.~(\ref{CreepDef}) given for this case are used in our model calculations and are shown in Table~\ref{TabSint}. The values of other parameters used in the calculation of sintering are also given in Table~\ref{TabSint}.

%********************************************************************

\printcredits

%********************************************************************

\bibliographystyle{cas-model2-names}

%\bibliography{litbib}

\begin{thebibliography}{109}
\expandafter\ifx\csname natexlab\endcsname\relax\def\natexlab#1{#1}\fi
\providecommand{\url}[1]{\texttt{#1}}
\providecommand{\href}[2]{#2}
\providecommand{\path}[1]{#1}
\providecommand{\DOIprefix}{doi:}
\providecommand{\ArXivprefix}{arXiv:}
\providecommand{\URLprefix}{URL: }
\providecommand{\Pubmedprefix}{pmid:}
\providecommand{\doi}[1]{\href{http://dx.doi.org/#1}{\path{#1}}}
\providecommand{\Pubmed}[1]{\href{pmid:#1}{\path{#1}}}
\providecommand{\bibinfo}[2]{#2}
\ifx\xfnm\relax \def\xfnm[#1]{\unskip,\space#1}\fi
%Type = Article
\bibitem[{{Bennett} and {McSween}(1996)}]{Ben96}
\bibinfo{author}{{Bennett}, Marvin~E., I.}, \bibinfo{author}{{McSween},
  Harry~Y., J.}, \bibinfo{year}{1996}.
\newblock \bibinfo{title}{{Revised model calculations for the thermal histories
  of ordinary chondrite parent bodies}}.
\newblock \bibinfo{journal}{Meteoritics and Planetary Science}
  \bibinfo{volume}{31}, \bibinfo{pages}{783--792}.
\newblock \DOIprefix\doi{10.1111/j.1945-5100.1996.tb02113.x}.
%Type = Inproceedings
\bibitem[{{Binzel} et~al.(2014){Binzel}, {DeMeo}, {Burt}, {Polishook},
  {Burbine}, {Bus}, {Tokunaga} and {Birlan}}]{Bin14}
\bibinfo{author}{{Binzel}, R.P.}, \bibinfo{author}{{DeMeo}, F.E.},
  \bibinfo{author}{{Burt}, B.J.}, \bibinfo{author}{{Polishook}, D.},
  \bibinfo{author}{{Burbine}, T.H.}, \bibinfo{author}{{Bus}, S.J.},
  \bibinfo{author}{{Tokunaga}, A.}, \bibinfo{author}{{Birlan}, M.},
  \bibinfo{year}{2014}.
\newblock \bibinfo{title}{{Meteorite Source Regions as Revealed by the
  Near-Earth Object Population}}, in: \bibinfo{booktitle}{AAS/Division for
  Planetary Sciences Meeting Abstracts \#46}, p. \bibinfo{pages}{213.06}.
%Type = Article
\bibitem[{{Birnstiel} et~al.(2016){Birnstiel}, {Fang} and {Johansen}}]{Bir16}
\bibinfo{author}{{Birnstiel}, T.}, \bibinfo{author}{{Fang}, M.},
  \bibinfo{author}{{Johansen}, A.}, \bibinfo{year}{2016}.
\newblock \bibinfo{title}{{Dust Evolution and the Formation of Planetesimals}}.
\newblock \bibinfo{journal}{\ssr} \bibinfo{volume}{205},
  \bibinfo{pages}{41--75}.
\newblock \DOIprefix\doi{10.1007/s11214-016-0256-1},
  \href{http://arxiv.org/abs/1604.02952}{\tt arXiv:1604.02952}.
%Type = Inproceedings
\bibitem[{{Boesenberg} et~al.(2014){Boesenberg}, {Weisberg}, {Greenwood},
  {Gibson} and {Franchi}}]{Boe14}
\bibinfo{author}{{Boesenberg}, J.S.}, \bibinfo{author}{{Weisberg}, M.K.},
  \bibinfo{author}{{Greenwood}, R.C.}, \bibinfo{author}{{Gibson}, J.M.},
  \bibinfo{author}{{Franchi}, I.A.}, \bibinfo{year}{2014}.
\newblock \bibinfo{title}{{The Anomalous Enstatite Meteorites {\textemdash}
  Part 2: The Recrystallized EL Meteorites}}, in: \bibinfo{booktitle}{Lunar and
  Planetary Science Conference}, p. \bibinfo{pages}{1486}.
%Type = Article
\bibitem[{{Bogard} et~al.(2010){Bogard}, {Dixon} and {Garrison}}]{Bog10}
\bibinfo{author}{{Bogard}, D.D.}, \bibinfo{author}{{Dixon}, E.T.},
  \bibinfo{author}{{Garrison}, D.H.}, \bibinfo{year}{2010}.
\newblock \bibinfo{title}{{Ar-Ar ages and thermal histories of enstatite
  meteorites}}.
\newblock \bibinfo{journal}{Meteoritics and Planetary Science}
  \bibinfo{volume}{45}, \bibinfo{pages}{723--742}.
\newblock \DOIprefix\doi{10.1111/j.1945-5100.2010.01060.x}.
%Type = Article
\bibitem[{{Brazzle} et~al.(1999){Brazzle}, {Pravdivtseva}, {Meshik} and
  {Hohenberg}}]{Bra99}
\bibinfo{author}{{Brazzle}, R.H.}, \bibinfo{author}{{Pravdivtseva}, O.V.},
  \bibinfo{author}{{Meshik}, A.P.}, \bibinfo{author}{{Hohenberg}, C.M.},
  \bibinfo{year}{1999}.
\newblock \bibinfo{title}{{Verification and interpretation of the I-Xe
  chronometer}}.
\newblock \bibinfo{journal}{\gca} \bibinfo{volume}{63},
  \bibinfo{pages}{739--760}.
\newblock \DOIprefix\doi{10.1016/S0016-7037(98)00314-7}.
%Type = Article
\bibitem[{{Britt} and {Consolmagno}(2003)}]{Bri03}
\bibinfo{author}{{Britt}, D.T.}, \bibinfo{author}{{Consolmagno}, G.J.},
  \bibinfo{year}{2003}.
\newblock \bibinfo{title}{{Stony meteorite porosities and densities: A review
  of the data through 2001}}.
\newblock \bibinfo{journal}{Meteoritics and Planetary Science}
  \bibinfo{volume}{38}, \bibinfo{pages}{1161--1180}.
\newblock \DOIprefix\doi{10.1111/j.1945-5100.2003.tb00305.x}.
%Type = Article
\bibitem[{{B{\"u}rgmann} and {Dresen}(2008)}]{Bur08}
\bibinfo{author}{{B{\"u}rgmann}, R.}, \bibinfo{author}{{Dresen}, G.},
  \bibinfo{year}{2008}.
\newblock \bibinfo{title}{{Rheology of the Lower Crust and Upper Mantle:
  Evidence from Rock Mechanics, Geodesy, and Field Observations}}.
\newblock \bibinfo{journal}{Annual Review of Earth and Planetary Sciences}
  \bibinfo{volume}{36}, \bibinfo{pages}{531--567}.
\newblock \DOIprefix\doi{10.1146/annurev.earth.36.031207.124326}.
%Type = Article
\bibitem[{{Burkhardt} et~al.(2008){Burkhardt}, {Kleine}, {Bourdon}, {Palme},
  {Zipfel}, {Friedrich} and {Ebel}}]{Kle08}
\bibinfo{author}{{Burkhardt}, C.}, \bibinfo{author}{{Kleine}, T.},
  \bibinfo{author}{{Bourdon}, B.}, \bibinfo{author}{{Palme}, H.},
  \bibinfo{author}{{Zipfel}, J.}, \bibinfo{author}{{Friedrich}, J.M.},
  \bibinfo{author}{{Ebel}, D.S.}, \bibinfo{year}{2008}.
\newblock \bibinfo{title}{{Hf W mineral isochron for Ca,Al-rich inclusions: Age
  of the solar system and the timing of core formation in planetesimals}}.
\newblock \bibinfo{journal}{\gca} \bibinfo{volume}{72},
  \bibinfo{pages}{6177--6197}.
\newblock \DOIprefix\doi{10.1016/j.gca.2008.10.023}.
%Type = Article
\bibitem[{{Busfield} et~al.(2008){Busfield}, {Turner} and {Gilmour}}]{Bus08}
\bibinfo{author}{{Busfield}, A.}, \bibinfo{author}{{Turner}, G.},
  \bibinfo{author}{{Gilmour}, J.D.}, \bibinfo{year}{2008}.
\newblock \bibinfo{title}{{Testing an integrated chronology: I-Xe analysis of
  enstatite meteorites and a eucrite}}.
\newblock \bibinfo{journal}{\maps} \bibinfo{volume}{43},
  \bibinfo{pages}{883--897}.
\newblock \DOIprefix\doi{10.1111/j.1945-5100.2008.tb01087.x}.
%Type = Article
\bibitem[{Bystricky et~al.(2016)Bystricky, Lawlis, Mackwell, Heidelbach and
  Raterron}]{Bys16}
\bibinfo{author}{Bystricky, M.}, \bibinfo{author}{Lawlis, J.},
  \bibinfo{author}{Mackwell, S.}, \bibinfo{author}{Heidelbach, F.},
  \bibinfo{author}{Raterron, P.}, \bibinfo{year}{2016}.
\newblock \bibinfo{title}{High-temperature deformation of enstatite
  aggregates}.
\newblock \bibinfo{journal}{Journal of Geophysical Research}
  \bibinfo{volume}{121}, \bibinfo{pages}{6384--6400}.
\newblock \DOIprefix\doi{10.1002/2016JB013011}.
%Type = Article
\bibitem[{{Callaway}(1959)}]{Cal59}
\bibinfo{author}{{Callaway}, J.}, \bibinfo{year}{1959}.
\newblock \bibinfo{title}{{Model for Lattice Thermal Conductivity at Low
  Temperatures}}.
\newblock \bibinfo{journal}{Physical Review} \bibinfo{volume}{113},
  \bibinfo{pages}{1046--1051}.
\newblock \DOIprefix\doi{10.1103/PhysRev.113.1046}.
%Type = Book
\bibitem[{{Carslaw} and {Jaeger}(1959)}]{Car59}
\bibinfo{author}{{Carslaw}, H.S.}, \bibinfo{author}{{Jaeger}, J.C.},
  \bibinfo{year}{1959}.
\newblock \bibinfo{title}{{Conduction of heat in solids}}.
%Type = Article
\bibitem[{{Cherniak} and {Van Orman}(2014)}]{Che14}
\bibinfo{author}{{Cherniak}, D.J.}, \bibinfo{author}{{Van Orman}, J.A.},
  \bibinfo{year}{2014}.
\newblock \bibinfo{title}{{Tungsten diffusion in olivine}}.
\newblock \bibinfo{journal}{\gca} \bibinfo{volume}{129},
  \bibinfo{pages}{1--12}.
\newblock \DOIprefix\doi{10.1016/j.gca.2013.12.020}.
%Type = Article
\bibitem[{{Collinet} and {Grove}(2020)}]{Col20}
\bibinfo{author}{{Collinet}, M.}, \bibinfo{author}{{Grove}, T.L.},
  \bibinfo{year}{2020}.
\newblock \bibinfo{title}{{Formation of primitive achondrites by partial
  melting of alkali-undepleted planetesimals in the inner solar system}}.
\newblock \bibinfo{journal}{\gca} \bibinfo{volume}{277},
  \bibinfo{pages}{358--376}.
\newblock \DOIprefix\doi{10.1016/j.gca.2020.03.004}.
%Type = Article
\bibitem[{{Crabb} and {Anders}(1981)}]{Cra81}
\bibinfo{author}{{Crabb}, J.}, \bibinfo{author}{{Anders}, E.},
  \bibinfo{year}{1981}.
\newblock \bibinfo{title}{{Noble gases in E-chondrites}}.
\newblock \bibinfo{journal}{\gca} \bibinfo{volume}{45},
  \bibinfo{pages}{2443--2464}.
\newblock \DOIprefix\doi{10.1016/0016-7037(81)90097-1}.
%Type = Article
\bibitem[{{Cuzzi} et~al.(2008){Cuzzi}, {Hogan} and {Shariff}}]{Cuz08}
\bibinfo{author}{{Cuzzi}, J.N.}, \bibinfo{author}{{Hogan}, R.C.},
  \bibinfo{author}{{Shariff}, K.}, \bibinfo{year}{2008}.
\newblock \bibinfo{title}{{Toward Planetesimals: Dense Chondrule Clumps in the
  Protoplanetary Nebula}}.
\newblock \bibinfo{journal}{\apj} \bibinfo{volume}{687},
  \bibinfo{pages}{1432--1447}.
\newblock \DOIprefix\doi{10.1086/591239},
  \href{http://arxiv.org/abs/0804.3526}{\tt arXiv:0804.3526}.
%Type = Article
\bibitem[{{D'Antona} and {Mazzitelli}(1994)}]{Dan94}
\bibinfo{author}{{D'Antona}, F.}, \bibinfo{author}{{Mazzitelli}, I.},
  \bibinfo{year}{1994}.
\newblock \bibinfo{title}{{New Pre--Main-Sequence Tracks for M <=2.5 M$_{sun}$
  as Tests of Opacities and Convection Model}}.
\newblock \bibinfo{journal}{\apjs} \bibinfo{volume}{90}, \bibinfo{pages}{467}.
\newblock \DOIprefix\doi{10.1086/191867}.
%Type = Article
\bibitem[{{Dauphas}(2017)}]{Dau17}
\bibinfo{author}{{Dauphas}, N.}, \bibinfo{year}{2017}.
\newblock \bibinfo{title}{{The isotopic nature of the Earth{\textquoteright}s
  accreting material through time}}.
\newblock \bibinfo{journal}{\nat} \bibinfo{volume}{541},
  \bibinfo{pages}{521--524}.
\newblock \DOIprefix\doi{10.1038/nature20830}.
%Type = Article
\bibitem[{Dodson(1973)}]{Dod73}
\bibinfo{author}{Dodson, M.H.}, \bibinfo{year}{1973}.
\newblock \bibinfo{title}{Clsoure temperature in cooling geochronlogical and
  petrological systems}.
\newblock \bibinfo{journal}{Contrib. Mineral. Petrol.} \bibinfo{volume}{40},
  \bibinfo{pages}{259--274}.
\newblock \DOIprefix\doi{10.1007/BF00373790}.
%Type = Article
\bibitem[{{El Goresy} et~al.(2017){El Goresy}, {Lin}, {Miyahara}, {Gannoun},
  {Boyet}, {Ohtani}, {Gillet}, {Trieloff}, {Simionovici}, {Feng} and
  {Lemelle}}]{EGo17}
\bibinfo{author}{{El Goresy}, A.}, \bibinfo{author}{{Lin}, Y.},
  \bibinfo{author}{{Miyahara}, M.}, \bibinfo{author}{{Gannoun}, A.},
  \bibinfo{author}{{Boyet}, M.}, \bibinfo{author}{{Ohtani}, E.},
  \bibinfo{author}{{Gillet}, P.}, \bibinfo{author}{{Trieloff}, M.},
  \bibinfo{author}{{Simionovici}, A.}, \bibinfo{author}{{Feng}, L.},
  \bibinfo{author}{{Lemelle}, L.}, \bibinfo{year}{2017}.
\newblock \bibinfo{title}{{Origin of EL3 chondrites: Evidence for variable C/O
  ratios during their course of formation{\textemdash}A state of the art
  scrutiny}}.
\newblock \bibinfo{journal}{Meteoritics and Planetary Science}
  \bibinfo{volume}{52}, \bibinfo{pages}{781--806}.
\newblock \DOIprefix\doi{10.1111/maps.12832}.
%Type = Article
\bibitem[{{Eugster} and {Lorenzetti}(2005)}]{Eug05}
\bibinfo{author}{{Eugster}, O.}, \bibinfo{author}{{Lorenzetti}, S.},
  \bibinfo{year}{2005}.
\newblock \bibinfo{title}{{Cosmic-ray exposure ages of four acapulcoites and
  two differentiated achondrites and evidence for a two-layer structure of the
  acapulcoite/lodranite parent asteroid}}.
\newblock \bibinfo{journal}{\gca} \bibinfo{volume}{69},
  \bibinfo{pages}{2675--2685}.
\newblock \DOIprefix\doi{10.1016/j.gca.2004.12.006}.
%Type = Article
\bibitem[{Fei et~al.(1997)Fei, Bertka and Finger}]{Fei97}
\bibinfo{author}{Fei, Y.}, \bibinfo{author}{Bertka, C.M.},
  \bibinfo{author}{Finger, L.W.}, \bibinfo{year}{1997}.
\newblock \bibinfo{title}{{High-Pressure Iron-Sulfur Compound, Fe3S2, and
  Melting Relations in the Fe-FeS System.}}
\newblock \bibinfo{journal}{Science} \bibinfo{volume}{275},
  \bibinfo{pages}{1621--1623}.
\newblock \DOIprefix\doi{10.1126/science.275.5306.1621}.
%Type = Article
\bibitem[{{Friedrich} et~al.(2015){Friedrich}, {Weisberg}, {Ebel}, {Biltz},
  {Corbett}, {Iotzov}, {Khan} and {Wolman}}]{Fri15}
\bibinfo{author}{{Friedrich}, J.M.}, \bibinfo{author}{{Weisberg}, M.K.},
  \bibinfo{author}{{Ebel}, D.S.}, \bibinfo{author}{{Biltz}, A.E.},
  \bibinfo{author}{{Corbett}, B.M.}, \bibinfo{author}{{Iotzov}, I.V.},
  \bibinfo{author}{{Khan}, W.S.}, \bibinfo{author}{{Wolman}, M.D.},
  \bibinfo{year}{2015}.
\newblock \bibinfo{title}{{Chondrule size and related physical properties: A
  compilation and evaluation of current data across all meteorite groups}}.
\newblock \bibinfo{journal}{Chemie der Erde / Geochemistry}
  \bibinfo{volume}{75}, \bibinfo{pages}{419--443}.
\newblock \DOIprefix\doi{10.1016/j.chemer.2014.08.003},
  \href{http://arxiv.org/abs/1408.6581}{\tt arXiv:1408.6581}.
%Type = Article
\bibitem[{{Gail} et~al.(2015){Gail}, {Henke} and {Trieloff}}]{Gai15}
\bibinfo{author}{{Gail}, H.P.}, \bibinfo{author}{{Henke}, S.},
  \bibinfo{author}{{Trieloff}, M.}, \bibinfo{year}{2015}.
\newblock \bibinfo{title}{{Thermal evolution and sintering of chondritic
  planetesimals. II. Improved treatment of the compaction process}}.
\newblock \bibinfo{journal}{\aap} \bibinfo{volume}{576}, \bibinfo{pages}{A60}.
\newblock \DOIprefix\doi{10.1051/0004-6361/201424278},
  \href{http://arxiv.org/abs/1411.2850}{\tt arXiv:1411.2850}.
%Type = Article
\bibitem[{{Gail} and {Trieloff}(2018)}]{Gai18}
\bibinfo{author}{{Gail}, H.P.}, \bibinfo{author}{{Trieloff}, M.},
  \bibinfo{year}{2018}.
\newblock \bibinfo{title}{{Thermal evolution and sintering of chondritic
  planetesimals. IV. Temperature dependence of heat conductivity of asteroids
  and meteorites}}.
\newblock \bibinfo{journal}{\aap} \bibinfo{volume}{615}, \bibinfo{pages}{A147}.
\newblock \DOIprefix\doi{10.1051/0004-6361/201732456},
  \href{http://arxiv.org/abs/1804.00574}{\tt arXiv:1804.00574}.
%Type = Article
\bibitem[{{Gail} and {Trieloff}(2019)}]{Gai19}
\bibinfo{author}{{Gail}, H.P.}, \bibinfo{author}{{Trieloff}, M.},
  \bibinfo{year}{2019}.
\newblock \bibinfo{title}{{Thermal history modelling of the L chondrite parent
  body}}.
\newblock \bibinfo{journal}{\aap} \bibinfo{volume}{628}, \bibinfo{pages}{A77}.
\newblock \DOIprefix\doi{10.1051/0004-6361/201936020},
  \href{http://arxiv.org/abs/1907.00805}{\tt arXiv:1907.00805}.
%Type = Article
\bibitem[{{Ganguly} et~al.(2007){Ganguly}, {Ito} and {Zhang}}]{Gan07}
\bibinfo{author}{{Ganguly}, J.}, \bibinfo{author}{{Ito}, M.},
  \bibinfo{author}{{Zhang}, X.}, \bibinfo{year}{2007}.
\newblock \bibinfo{title}{{Cr diffusion in orthopyroxene: Experimental
  determination, $^{53}$Mn- $^{53}$Cr thermochronology, and planetary
  applications}}.
\newblock \bibinfo{journal}{\gca} \bibinfo{volume}{71},
  \bibinfo{pages}{3915--3925}.
\newblock \DOIprefix\doi{10.1016/j.gca.2007.05.023}.
%Type = Article
\bibitem[{{G\"opel} et~al.(1994){G\"opel}, {Manhes} and {Allegre}}]{Goe94}
\bibinfo{author}{{G\"opel}, C.}, \bibinfo{author}{{Manhes}, G.},
  \bibinfo{author}{{Allegre}, C.J.}, \bibinfo{year}{1994}.
\newblock \bibinfo{title}{{U-Pb systematics of phosphates from equilibrated
  ordinary chondrites}}.
\newblock \bibinfo{journal}{Earth and Planetary Science Letters}
  \bibinfo{volume}{121}, \bibinfo{pages}{153--171}.
\newblock \DOIprefix\doi{10.1016/0012-821X(94)90038-8}.
%Type = Article
\bibitem[{{Haisch} et~al.(2001){Haisch}, {Lada} and {Lada}}]{Hai01}
\bibinfo{author}{{Haisch}, Karl~E., J.}, \bibinfo{author}{{Lada}, E.A.},
  \bibinfo{author}{{Lada}, C.J.}, \bibinfo{year}{2001}.
\newblock \bibinfo{title}{{Disk Frequencies and Lifetimes in Young Clusters}}.
\newblock \bibinfo{journal}{\apjl} \bibinfo{volume}{553},
  \bibinfo{pages}{L153--L156}.
\newblock \DOIprefix\doi{10.1086/320685},
  \href{http://arxiv.org/abs/astro-ph/0104347}{\tt arXiv:astro-ph/0104347}.
%Type = Article
\bibitem[{{Harrison} and {Grimm}(2010)}]{Har10}
\bibinfo{author}{{Harrison}, K.P.}, \bibinfo{author}{{Grimm}, R.E.},
  \bibinfo{year}{2010}.
\newblock \bibinfo{title}{{Thermal constraints on the early history of the
  H-chondrite parent body reconsidered}}.
\newblock \bibinfo{journal}{\gca} \bibinfo{volume}{74},
  \bibinfo{pages}{5410--5423}.
\newblock \DOIprefix\doi{10.1016/j.gca.2010.05.034}.
%Type = Article
\bibitem[{{Henke} et~al.(2016){Henke}, {Gail} and {Trieloff}}]{Hen16}
\bibinfo{author}{{Henke}, S.}, \bibinfo{author}{{Gail}, H.P.},
  \bibinfo{author}{{Trieloff}, M.}, \bibinfo{year}{2016}.
\newblock \bibinfo{title}{{Thermal evolution and sintering of chondritic
  planetesimals. III. Modelling the heat conductivity of porous chondrite
  material}}.
\newblock \bibinfo{journal}{\aap} \bibinfo{volume}{589}, \bibinfo{pages}{A41}.
\newblock \DOIprefix\doi{10.1051/0004-6361/201527687},
  \href{http://arxiv.org/abs/1602.00292}{\tt arXiv:1602.00292}.
%Type = Article
\bibitem[{{Henke} et~al.(2013){Henke}, {Gail}, {Trieloff} and
  {Schwarz}}]{Hen13}
\bibinfo{author}{{Henke}, S.}, \bibinfo{author}{{Gail}, H.P.},
  \bibinfo{author}{{Trieloff}, M.}, \bibinfo{author}{{Schwarz}, W.H.},
  \bibinfo{year}{2013}.
\newblock \bibinfo{title}{{Thermal evolution model for the H chondrite
  asteroid-instantaneous formation versus protracted accretion}}.
\newblock \bibinfo{journal}{\icarus} \bibinfo{volume}{226},
  \bibinfo{pages}{212--228}.
\newblock \DOIprefix\doi{10.1016/j.icarus.2013.05.034}.
%Type = Article
\bibitem[{{Henke} et~al.(2012a){Henke}, {Gail}, {Trieloff}, {Schwarz} and
  {Kleine}}]{Gai12}
\bibinfo{author}{{Henke}, S.}, \bibinfo{author}{{Gail}, H.P.},
  \bibinfo{author}{{Trieloff}, M.}, \bibinfo{author}{{Schwarz}, W.H.},
  \bibinfo{author}{{Kleine}, T.}, \bibinfo{year}{2012}a.
\newblock \bibinfo{title}{{Thermal evolution and sintering of chondritic
  planetesimals}}.
\newblock \bibinfo{journal}{\aap} \bibinfo{volume}{537}, \bibinfo{pages}{A45}.
\newblock \DOIprefix\doi{10.1051/0004-6361/201117177},
  \href{http://arxiv.org/abs/1110.4818}{\tt arXiv:1110.4818}.
%Type = Article
\bibitem[{{Henke} et~al.(2012b){Henke}, {Gail}, {Trieloff}, {Schwarz} and
  {Kleine}}]{Hen12}
\bibinfo{author}{{Henke}, S.}, \bibinfo{author}{{Gail}, H.P.},
  \bibinfo{author}{{Trieloff}, M.}, \bibinfo{author}{{Schwarz}, W.H.},
  \bibinfo{author}{{Kleine}, T.}, \bibinfo{year}{2012}b.
\newblock \bibinfo{title}{{Thermal history modelling of the H chondrite parent
  body}}.
\newblock \bibinfo{journal}{\aap} \bibinfo{volume}{545}, \bibinfo{pages}{A135}.
\newblock \DOIprefix\doi{10.1051/0004-6361/201219100},
  \href{http://arxiv.org/abs/1208.4633}{\tt arXiv:1208.4633}.
%Type = Article
\bibitem[{{Hevey} and {Sanders}(2006)}]{Hev06}
\bibinfo{author}{{Hevey}, P.J.}, \bibinfo{author}{{Sanders}, I.S.},
  \bibinfo{year}{2006}.
\newblock \bibinfo{title}{{A model for planetesimal meltdown by $^{26}$Al and
  its implications for meteorite parent bodies}}.
\newblock \bibinfo{journal}{Meteoritics and Planetary Science}
  \bibinfo{volume}{41}, \bibinfo{pages}{95--106}.
\newblock \DOIprefix\doi{10.1111/j.1945-5100.2006.tb00195.x}.
%Type = Article
\bibitem[{{Hopp} et~al.(2016){Hopp}, {Trieloff} and {Ott}}]{Hop16}
\bibinfo{author}{{Hopp}, J.}, \bibinfo{author}{{Trieloff}, M.},
  \bibinfo{author}{{Ott}, U.}, \bibinfo{year}{2016}.
\newblock \bibinfo{title}{{I-Xe ages of enstatite chondrites}}.
\newblock \bibinfo{journal}{\gca} \bibinfo{volume}{174},
  \bibinfo{pages}{196--210}.
\newblock \DOIprefix\doi{10.1016/j.gca.2015.11.014}.
%Type = Article
\bibitem[{{Hopp} et~al.(2014){Hopp}, {Trieloff}, {Ott}, {Korochantseva} and
  {Buykin}}]{Hop14}
\bibinfo{author}{{Hopp}, J.}, \bibinfo{author}{{Trieloff}, M.},
  \bibinfo{author}{{Ott}, U.}, \bibinfo{author}{{Korochantseva}, E.V.},
  \bibinfo{author}{{Buykin}, A.I.}, \bibinfo{year}{2014}.
\newblock \bibinfo{title}{{$^{39}$Ar-$^{40}$Ar chronology of the enstatite
  chondrite parent bodies}}.
\newblock \bibinfo{journal}{Meteoritics and Planetary Science}
  \bibinfo{volume}{49}, \bibinfo{pages}{358--372}.
\newblock \DOIprefix\doi{10.1111/maps.12243}.
%Type = Article
\bibitem[{{Hoppe}(2009)}]{Hop09}
\bibinfo{author}{{Hoppe}, P.}, \bibinfo{year}{2009}.
\newblock \bibinfo{title}{{Meteorites}}.
\newblock \bibinfo{journal}{Landolt B\&ouml;rnstein} \bibinfo{volume}{4B},
  \bibinfo{pages}{582}.
\newblock \DOIprefix\doi{10.1007/978-3-540-88055-4_30}.
%Type = Inbook
\bibitem[{{Huss} et~al.(2006){Huss}, {Rubin} and {Grossman}}]{Hus06}
\bibinfo{author}{{Huss}, G.R.}, \bibinfo{author}{{Rubin}, A.E.},
  \bibinfo{author}{{Grossman}, J.N.}, \bibinfo{year}{2006}.
\newblock \bibinfo{title}{{Thermal Metamorphism in Chondrites}}.
\newblock p. \bibinfo{pages}{567}.
%Type = Article
\bibitem[{{Ito} and {Ganguly}(2006)}]{Ito06}
\bibinfo{author}{{Ito}, M.}, \bibinfo{author}{{Ganguly}, J.},
  \bibinfo{year}{2006}.
\newblock \bibinfo{title}{{Diffusion kinetics of Cr in olivine and $^{53}$Mn-
  $^{53}$Cr thermochronology of early solar system objects}}.
\newblock \bibinfo{journal}{\gca} \bibinfo{volume}{70},
  \bibinfo{pages}{799--809}.
\newblock \DOIprefix\doi{10.1016/j.gca.2005.09.020}.
%Type = Article
\bibitem[{{Jarosewich}(1990)}]{Jar90}
\bibinfo{author}{{Jarosewich}, E.}, \bibinfo{year}{1990}.
\newblock \bibinfo{title}{{Chemical Analyses of Meteorites: A Compilation of
  Stony and Iron Meteorite Analyses}}.
\newblock \bibinfo{journal}{Meteoritics} \bibinfo{volume}{25},
  \bibinfo{pages}{323}.
\newblock \DOIprefix\doi{10.1111/j.1945-5100.1990.tb00717.x}.
%Type = Article
\bibitem[{{Javoy} et~al.(2010){Javoy}, {Kaminski}, {Guyot}, {Andrault},
  {Sanloup}, {Moreira}, {Labrosse}, {Jambon}, {Agrinier}, {Davaille} and
  {Jaupart}}]{Jav10}
\bibinfo{author}{{Javoy}, M.}, \bibinfo{author}{{Kaminski}, E.},
  \bibinfo{author}{{Guyot}, F.}, \bibinfo{author}{{Andrault}, D.},
  \bibinfo{author}{{Sanloup}, C.}, \bibinfo{author}{{Moreira}, M.},
  \bibinfo{author}{{Labrosse}, S.}, \bibinfo{author}{{Jambon}, A.},
  \bibinfo{author}{{Agrinier}, P.}, \bibinfo{author}{{Davaille}, A.},
  \bibinfo{author}{{Jaupart}, C.}, \bibinfo{year}{2010}.
\newblock \bibinfo{title}{{The chemical composition of the Earth: Enstatite
  chondrite models}}.
\newblock \bibinfo{journal}{Earth and Planetary Science Letters}
  \bibinfo{volume}{293}, \bibinfo{pages}{259--268}.
\newblock \DOIprefix\doi{10.1016/j.epsl.2010.02.033}.
%Type = Article
\bibitem[{{Johansen} et~al.(2014){Johansen}, {Blum}, {Tanaka}, {Ormel},
  {Bizzarro} and {Rickman}}]{Joh14}
\bibinfo{author}{{Johansen}, A.}, \bibinfo{author}{{Blum}, J.},
  \bibinfo{author}{{Tanaka}, H.}, \bibinfo{author}{{Ormel}, C.},
  \bibinfo{author}{{Bizzarro}, M.}, \bibinfo{author}{{Rickman}, H.},
  \bibinfo{year}{2014}.
\newblock \bibinfo{title}{{The Multifaceted Planetesimal Formation Process}}.
\newblock \bibinfo{journal}{Protostars and Planets VI} ,
  \bibinfo{pages}{547--570}\DOIprefix\doi{10.2458/azu_uapress_9780816531240-ch024},
  \href{http://arxiv.org/abs/1402.1344}{\tt arXiv:1402.1344}.
%Type = Incollection
\bibitem[{{Johansen} et~al.(2015a){Johansen}, {Jacquet}, {Cuzzi}, {Morbidelli}
  and {Gounelle}}]{Joh15}
\bibinfo{author}{{Johansen}, A.}, \bibinfo{author}{{Jacquet}, E.},
  \bibinfo{author}{{Cuzzi}, J.N.}, \bibinfo{author}{{Morbidelli}, A.},
  \bibinfo{author}{{Gounelle}, M.}, \bibinfo{year}{2015}a.
\newblock \bibinfo{title}{{New Paradigms For Asteroid Formation.}}, in:
  \bibinfo{editor}{Michel, P.}, \bibinfo{editor}{DeMeo, F.E.},
  \bibinfo{editor}{{Bottke Jr.}, W.F.} (Eds.), \bibinfo{booktitle}{Asteroids
  IV}. \bibinfo{publisher}{University of Arizona press},
  \bibinfo{address}{Tucson}, pp. \bibinfo{pages}{471--492}.
\newblock \DOIprefix\doi{10.2458/azu_uapress_9780816532131-ch025}.
%Type = Article
\bibitem[{{Johansen} et~al.(2015b){Johansen}, {Mac Low}, {Lacerda} and
  {Bizzarro}}]{Joh15a}
\bibinfo{author}{{Johansen}, A.}, \bibinfo{author}{{Mac Low}, M.M.},
  \bibinfo{author}{{Lacerda}, P.}, \bibinfo{author}{{Bizzarro}, M.},
  \bibinfo{year}{2015}b.
\newblock \bibinfo{title}{{Growth of asteroids, planetary embryos, and Kuiper
  belt objects by chondrule accretion}}.
\newblock \bibinfo{journal}{Science Advances} \bibinfo{volume}{1},
  \bibinfo{pages}{1500109}.
\newblock \DOIprefix\doi{10.1126/sciadv.1500109},
  \href{http://arxiv.org/abs/1503.07347}{\tt arXiv:1503.07347}.
%Type = Article
\bibitem[{{Johansen} et~al.(2007){Johansen}, {Oishi}, {Mac Low}, {Klahr},
  {Henning} and {Youdin}}]{Joh07}
\bibinfo{author}{{Johansen}, A.}, \bibinfo{author}{{Oishi}, J.S.},
  \bibinfo{author}{{Mac Low}, M.M.}, \bibinfo{author}{{Klahr}, H.},
  \bibinfo{author}{{Henning}, T.}, \bibinfo{author}{{Youdin}, A.},
  \bibinfo{year}{2007}.
\newblock \bibinfo{title}{{Rapid planetesimal formation in turbulent
  circumstellar disks}}.
\newblock \bibinfo{journal}{\nat} \bibinfo{volume}{448},
  \bibinfo{pages}{1022--1025}.
\newblock \DOIprefix\doi{10.1038/nature06086},
  \href{http://arxiv.org/abs/0708.3890}{\tt arXiv:0708.3890}.
%Type = Inbook
\bibitem[{Karato(2013)}]{Kar13}
\bibinfo{author}{Karato, S.I.}, \bibinfo{year}{2013}.
\newblock \bibinfo{title}{Rheological Properties of Minerals and Rocks}.
  \bibinfo{publisher}{John Wiley \& Sons, Ltd}. chapter~\bibinfo{chapter}{4}.
\newblock pp. \bibinfo{pages}{94--144}.
\newblock \DOIprefix\doi{https://doi.org/10.1002/9781118529492.ch4}.
%Type = Article
\bibitem[{{Keil}(1989)}]{Kei89}
\bibinfo{author}{{Keil}, K.}, \bibinfo{year}{1989}.
\newblock \bibinfo{title}{{Enstatite Meteorites and Their Parent Bodies}}.
\newblock \bibinfo{journal}{Meteoritics} \bibinfo{volume}{24},
  \bibinfo{pages}{195}.
\newblock \DOIprefix\doi{10.1111/j.1945-5100.1989.tb00694.x}.
%Type = Phdthesis
\bibitem[{{Kennedy}(1981)}]{Ken81}
\bibinfo{author}{{Kennedy}, B.M.}, \bibinfo{year}{1981}.
\newblock \bibinfo{title}{{Potassium-argon and iodine-xenon gas retention ages
  of enstatite chondrite meteorites}}.
\newblock Ph.D. thesis. Washington Univ., Saint Louis, MO.
%Type = Article
\bibitem[{{Kennedy} et~al.(1988){Kennedy}, {Hudson}, {Hohenberg} and
  {Podosek}}]{Ken88}
\bibinfo{author}{{Kennedy}, B.M.}, \bibinfo{author}{{Hudson}, B.},
  \bibinfo{author}{{Hohenberg}, C.M.}, \bibinfo{author}{{Podosek}, F.A.},
  \bibinfo{year}{1988}.
\newblock \bibinfo{title}{{$^{129}$I/ $^{127}$I variations among enstatite
  chondrites}}.
\newblock \bibinfo{journal}{\gca} \bibinfo{volume}{52},
  \bibinfo{pages}{101--111}.
\newblock \DOIprefix\doi{10.1016/0016-7037(88)90059-2}.
%Type = Article
\bibitem[{{Klahr} and {Schreiber}(2020)}]{Kla20}
\bibinfo{author}{{Klahr}, H.}, \bibinfo{author}{{Schreiber}, A.},
  \bibinfo{year}{2020}.
\newblock \bibinfo{title}{{Turbulence Sets the Length Scale for Planetesimal
  Formation: Local 2D Simulations of Streaming Instability and Planetesimal
  Formation}}.
\newblock \bibinfo{journal}{\apj} \bibinfo{volume}{901}, \bibinfo{pages}{54}.
\newblock \DOIprefix\doi{10.3847/1538-4357/abac58},
  \href{http://arxiv.org/abs/2007.10696}{\tt arXiv:2007.10696}.
%Type = Article
\bibitem[{{Kleine} et~al.(2005){Kleine}, {Mezger}, {Palme}, {Scherer} and
  {M{\"u}nker}}]{Kle05}
\bibinfo{author}{{Kleine}, T.}, \bibinfo{author}{{Mezger}, K.},
  \bibinfo{author}{{Palme}, H.}, \bibinfo{author}{{Scherer}, E.},
  \bibinfo{author}{{M{\"u}nker}, C.}, \bibinfo{year}{2005}.
\newblock \bibinfo{title}{{Early core formation in asteroids and late accretion
  of chondrite parent bodies: Evidence from $^{182}$Hf- $^{182}$W in CAIs,
  metal-rich chondrites, and iron meteorites}}.
\newblock \bibinfo{journal}{\gca} \bibinfo{volume}{69},
  \bibinfo{pages}{5805--5818}.
\newblock \DOIprefix\doi{10.1016/j.gca.2005.07.012}.
%Type = Article
\bibitem[{{Kleine} et~al.(2009){Kleine}, {Touboul}, {Bourdon}, {Nimmo},
  {Mezger}, {Palme}, {Jacobsen}, {Yin} and {Halliday}}]{Kle09}
\bibinfo{author}{{Kleine}, T.}, \bibinfo{author}{{Touboul}, M.},
  \bibinfo{author}{{Bourdon}, B.}, \bibinfo{author}{{Nimmo}, F.},
  \bibinfo{author}{{Mezger}, K.}, \bibinfo{author}{{Palme}, H.},
  \bibinfo{author}{{Jacobsen}, S.B.}, \bibinfo{author}{{Yin}, Q.Z.},
  \bibinfo{author}{{Halliday}, A.N.}, \bibinfo{year}{2009}.
\newblock \bibinfo{title}{{Hf-W chronology of the accretion and early evolution
  of asteroids and terrestrial planets}}.
\newblock \bibinfo{journal}{\gca} \bibinfo{volume}{73},
  \bibinfo{pages}{5150--5188}.
\newblock \DOIprefix\doi{10.1016/j.gca.2008.11.047}.
%Type = Article
\bibitem[{{Kleine} et~al.(2008){Kleine}, {Touboul}, {Van Orman}, {Bourdon},
  {Maden}, {Mezger} and {Halliday}}]{Kle08b}
\bibinfo{author}{{Kleine}, T.}, \bibinfo{author}{{Touboul}, M.},
  \bibinfo{author}{{Van Orman}, J.A.}, \bibinfo{author}{{Bourdon}, B.},
  \bibinfo{author}{{Maden}, C.}, \bibinfo{author}{{Mezger}, K.},
  \bibinfo{author}{{Halliday}, A.N.}, \bibinfo{year}{2008}.
\newblock \bibinfo{title}{{Hf-W thermochronometry: Closure temperature and
  constraints on the accretion and cooling history of the H chondrite parent
  body}}.
\newblock \bibinfo{journal}{Earth and Planetary Science Letters}
  \bibinfo{volume}{270}, \bibinfo{pages}{106--118}.
\newblock \DOIprefix\doi{10.1016/j.epsl.2008.03.013}.
%Type = Article
\bibitem[{{Korochantseva} et~al.(2007){Korochantseva}, {Trieloff}, {Lorenz},
  {Buykin}, {Ivanova}, {Schwarz}, {Hopp} and {Jessberger}}]{Kor07}
\bibinfo{author}{{Korochantseva}, E.V.}, \bibinfo{author}{{Trieloff}, M.},
  \bibinfo{author}{{Lorenz}, C.A.}, \bibinfo{author}{{Buykin}, A.I.},
  \bibinfo{author}{{Ivanova}, M.A.}, \bibinfo{author}{{Schwarz}, W.H.},
  \bibinfo{author}{{Hopp}, J.}, \bibinfo{author}{{Jessberger}, E.K.},
  \bibinfo{year}{2007}.
\newblock \bibinfo{title}{{L-chondrite asteroid breakup tied to Ordovician
  meteorite shower by multiple isochron 40Ar-39Ar dating}}.
\newblock \bibinfo{journal}{\maps} \bibinfo{volume}{42},
  \bibinfo{pages}{113--130}.
\newblock \DOIprefix\doi{10.1111/j.1945-5100.2007.tb00221.x}.
%Type = Inbook
\bibitem[{{Krot} et~al.(2005){Krot}, {Keil}, {Goodrich}, {Scott} and
  {Weisberg}}]{Krot05}
\bibinfo{author}{{Krot}, A.N.}, \bibinfo{author}{{Keil}, K.},
  \bibinfo{author}{{Goodrich}, C.A.}, \bibinfo{author}{{Scott}, E.R.D.},
  \bibinfo{author}{{Weisberg}, M.K.}, \bibinfo{year}{2005}.
\newblock \bibinfo{title}{{Classification of Meteorites}}.
  volume~\bibinfo{volume}{1}.
\newblock p.~\bibinfo{pages}{83}.
%Type = Inproceedings
\bibitem[{{La Tourrette} et~al.(1995){La Tourrette}, {Fahey} and
  {Wasserburg}}]{LaT98}
\bibinfo{author}{{La Tourrette}, T.Z.}, \bibinfo{author}{{Fahey}, A.J.},
  \bibinfo{author}{{Wasserburg}, G.J.}, \bibinfo{year}{1995}.
\newblock \bibinfo{title}{{The Effects of Ionic Radius and Charge on Diffusion
  of Mg, Ca, Ba, Ti, Zr, Nd, and Yb in Basaltic Melt}}, in:
  \bibinfo{booktitle}{Lunar and Planetary Science Conference}, p.
  \bibinfo{pages}{829}.
%Type = Article
\bibitem[{{Lee} and {Halliday}(2000)}]{Lee00}
\bibinfo{author}{{Lee}, D.C.}, \bibinfo{author}{{Halliday}, A.N.},
  \bibinfo{year}{2000}.
\newblock \bibinfo{title}{{Accretion of Primitive Planetesimals: Hf-W Isotopic
  Evidence from Enstatite Chondrites}}.
\newblock \bibinfo{journal}{Science} \bibinfo{volume}{288},
  \bibinfo{pages}{1629--1631}.
\newblock \DOIprefix\doi{10.1126/science.288.5471.1629}.
%Type = Article
\bibitem[{{Levison} et~al.(2015){Levison}, {Kretke} and {Duncan}}]{Lev15}
\bibinfo{author}{{Levison}, H.F.}, \bibinfo{author}{{Kretke}, K.A.},
  \bibinfo{author}{{Duncan}, M.J.}, \bibinfo{year}{2015}.
\newblock \bibinfo{title}{{Growing the gas-giant planets by the gradual
  accumulation of pebbles}}.
\newblock \bibinfo{journal}{\nat} \bibinfo{volume}{524},
  \bibinfo{pages}{322--324}.
\newblock \DOIprefix\doi{10.1038/nature14675},
  \href{http://arxiv.org/abs/1510.02094}{\tt arXiv:1510.02094}.
%Type = Article
\bibitem[{{Lucas} et~al.(2019){Lucas}, {Emery}, {MacLennan}, {Pinilla-Alonso},
  {Cartwright}, {Lindsay}, {Reddy}, {Sanchez}, {Thomas} and {Lorenzi}}]{Luc19}
\bibinfo{author}{{Lucas}, M.P.}, \bibinfo{author}{{Emery}, J.P.},
  \bibinfo{author}{{MacLennan}, E.M.}, \bibinfo{author}{{Pinilla-Alonso}, N.},
  \bibinfo{author}{{Cartwright}, R.J.}, \bibinfo{author}{{Lindsay}, S.S.},
  \bibinfo{author}{{Reddy}, V.}, \bibinfo{author}{{Sanchez}, J.A.},
  \bibinfo{author}{{Thomas}, C.A.}, \bibinfo{author}{{Lorenzi}, V.},
  \bibinfo{year}{2019}.
\newblock \bibinfo{title}{{Hungaria asteroid region telescopic spectral survey
  (HARTSS) II: Spectral homogeneity among Hungaria family asteroids}}.
\newblock \bibinfo{journal}{\icarus} \bibinfo{volume}{322},
  \bibinfo{pages}{227--250}.
\newblock \DOIprefix\doi{10.1016/j.icarus.2018.12.010}.
%Type = Article
\bibitem[{{Macke} et~al.(2010){Macke}, {Consolmagno}, {Britt} and
  {Hutson}}]{Mac10}
\bibinfo{author}{{Macke}, R.J.}, \bibinfo{author}{{Consolmagno}, G.J.},
  \bibinfo{author}{{Britt}, D.T.}, \bibinfo{author}{{Hutson}, M.L.},
  \bibinfo{year}{2010}.
\newblock \bibinfo{title}{{Enstatite chondrite density, magnetic
  susceptibility, and porosity}}.
\newblock \bibinfo{journal}{Meteoritics and Planetary Science}
  \bibinfo{volume}{45}, \bibinfo{pages}{1513--1526}.
\newblock \DOIprefix\doi{10.1111/j.1945-5100.2010.01129.x}.
%Type = Article
\bibitem[{{McCoy} et~al.(1999){McCoy}, {Dickinson} and {Lofgren}}]{McC99}
\bibinfo{author}{{McCoy}, T.J.}, \bibinfo{author}{{Dickinson}, T.L.},
  \bibinfo{author}{{Lofgren}, G.E.}, \bibinfo{year}{1999}.
\newblock \bibinfo{title}{{Partial melting of the Indarch (EH4) Meteorite: A
  textural, chemical and phase relations view of melting and melt migration}}.
\newblock \bibinfo{journal}{Meteoritics and Planetary Science}
  \bibinfo{volume}{34}, \bibinfo{pages}{735--746}.
\newblock \DOIprefix\doi{10.1111/j.1945-5100.1999.tb01386.x}.
%Type = Article
\bibitem[{{Miyamoto} et~al.(1981){Miyamoto}, {Fujii} and {Takeda}}]{Miy82}
\bibinfo{author}{{Miyamoto}, M.}, \bibinfo{author}{{Fujii}, N.},
  \bibinfo{author}{{Takeda}, H.}, \bibinfo{year}{1981}.
\newblock \bibinfo{title}{{Ordinary chondrite parent body - An internal heating
  model}}.
\newblock \bibinfo{journal}{Lunar and Planetary Science Conference Proceedings}
  \bibinfo{volume}{12}, \bibinfo{pages}{1145--1152}.
%Type = Article
\bibitem[{{Monnereau} et~al.(2013){Monnereau}, {Toplis}, {Baratoux} and
  {Guignard}}]{Mon13}
\bibinfo{author}{{Monnereau}, M.}, \bibinfo{author}{{Toplis}, M.J.},
  \bibinfo{author}{{Baratoux}, D.}, \bibinfo{author}{{Guignard}, J.},
  \bibinfo{year}{2013}.
\newblock \bibinfo{title}{{Thermal history of the H-chondrite parent body:
  Implications for metamorphic grade and accretionary time-scales}}.
\newblock \bibinfo{journal}{\gca} \bibinfo{volume}{119},
  \bibinfo{pages}{302--321}.
\newblock \DOIprefix\doi{10.1016/j.gca.2013.05.035}.
%Type = Article
\bibitem[{{Morbidelli} and {Raymond}(2016)}]{Mor16}
\bibinfo{author}{{Morbidelli}, A.}, \bibinfo{author}{{Raymond}, S.N.},
  \bibinfo{year}{2016}.
\newblock \bibinfo{title}{{Challenges in planet formation}}.
\newblock \bibinfo{journal}{Journal of Geophysical Research (Planets)}
  \bibinfo{volume}{121}, \bibinfo{pages}{1962--1980}.
\newblock \DOIprefix\doi{10.1002/2016JE005088},
  \href{http://arxiv.org/abs/1610.07202}{\tt arXiv:1610.07202}.
%Type = Article
\bibitem[{{Neumann} et~al.(2018){Neumann}, {Henke}, {Breuer}, {Gail},
  {Schwarz}, {Trieloff}, {Hopp} and {Spohn}}]{Neu18}
\bibinfo{author}{{Neumann}, W.}, \bibinfo{author}{{Henke}, S.},
  \bibinfo{author}{{Breuer}, D.}, \bibinfo{author}{{Gail}, H.P.},
  \bibinfo{author}{{Schwarz}, W.H.}, \bibinfo{author}{{Trieloff}, M.},
  \bibinfo{author}{{Hopp}, J.}, \bibinfo{author}{{Spohn}, T.},
  \bibinfo{year}{2018}.
\newblock \bibinfo{title}{{Modeling the evolution of the parent body of
  acapulcoites and lodranites: A case study for partially differentiated
  asteroids}}.
\newblock \bibinfo{journal}{\icarus} \bibinfo{volume}{311},
  \bibinfo{pages}{146--169}.
\newblock \DOIprefix\doi{10.1016/j.icarus.2018.03.024}.
%Type = Article
\bibitem[{{Opeil} et~al.(2012){Opeil}, {Consolmagno}, {Safarik} and
  {Britt}}]{Ope12}
\bibinfo{author}{{Opeil}, C.P.}, \bibinfo{author}{{Consolmagno}, G.J.},
  \bibinfo{author}{{Safarik}, D.J.}, \bibinfo{author}{{Britt}, D.T.},
  \bibinfo{year}{2012}.
\newblock \bibinfo{title}{{Stony meteorite thermal properties and their
  relationship with meteorite chemical and physical states}}.
\newblock \bibinfo{journal}{Meteoritics and Planetary Science}
  \bibinfo{volume}{47}, \bibinfo{pages}{319--329}.
\newblock \DOIprefix\doi{10.1111/j.1945-5100.2012.01331.x}.
%Type = Article
\bibitem[{{Pape} et~al.(2019){Pape}, {Mezger}, {Bouvier} and
  {Baumgartner}}]{Pap19}
\bibinfo{author}{{Pape}, J.}, \bibinfo{author}{{Mezger}, K.},
  \bibinfo{author}{{Bouvier}, A.S.}, \bibinfo{author}{{Baumgartner}, L.P.},
  \bibinfo{year}{2019}.
\newblock \bibinfo{title}{{Time and duration of chondrule formation:
  Constraints from $^{26}$Al-$^{26}$Mg ages of individual chondrules}}.
\newblock \bibinfo{journal}{\gca} \bibinfo{volume}{244},
  \bibinfo{pages}{416--436}.
\newblock \DOIprefix\doi{10.1016/j.gca.2018.10.017}.
%Type = Article
\bibitem[{{Patzer} and {Schultz}(2001)}]{Pat01}
\bibinfo{author}{{Patzer}, A.}, \bibinfo{author}{{Schultz}, L.},
  \bibinfo{year}{2001}.
\newblock \bibinfo{title}{{Noble gases in enstatite chondrites I: Exposure
  ages, pairing, and weathering effects}}.
\newblock \bibinfo{journal}{Meteoritics and Planetary Science}
  \bibinfo{volume}{36}, \bibinfo{pages}{947--961}.
\newblock \DOIprefix\doi{10.1111/j.1945-5100.2001.tb01932.x}.
%Type = Article
\bibitem[{{Pellas} et~al.(1997){Pellas}, {Fi{\'e}ni}, {Trieloff} and
  {Jessberger}}]{Pel97}
\bibinfo{author}{{Pellas}, P.}, \bibinfo{author}{{Fi{\'e}ni}, C.},
  \bibinfo{author}{{Trieloff}, M.}, \bibinfo{author}{{Jessberger}, E.K.},
  \bibinfo{year}{1997}.
\newblock \bibinfo{title}{{The cooling history of the Acapulco meteorite as
  recorded by the $^{244}$Pu and $^{40}$Ar- $^{39}$Ar chronometers}}.
\newblock \bibinfo{journal}{\gca} \bibinfo{volume}{61},
  \bibinfo{pages}{3477--3501}.
\newblock \DOIprefix\doi{10.1016/S0016-7037(97)00167-1}.
%Type = Article
\bibitem[{{Pfalzner} et~al.(2015){Pfalzner}, {Davies}, {Gounelle}, {Johansen},
  {M{\"u}nker}, {Lacerda}, {Portegies Zwart}, {Testi}, {Trieloff} and
  {Veras}}]{Pfa15}
\bibinfo{author}{{Pfalzner}, S.}, \bibinfo{author}{{Davies}, M.B.},
  \bibinfo{author}{{Gounelle}, M.}, \bibinfo{author}{{Johansen}, A.},
  \bibinfo{author}{{M{\"u}nker}, C.}, \bibinfo{author}{{Lacerda}, P.},
  \bibinfo{author}{{Portegies Zwart}, S.}, \bibinfo{author}{{Testi}, L.},
  \bibinfo{author}{{Trieloff}, M.}, \bibinfo{author}{{Veras}, D.},
  \bibinfo{year}{2015}.
\newblock \bibinfo{title}{{The formation of the solar system}}.
\newblock \bibinfo{journal}{Physica Scripta} \bibinfo{volume}{90},
  \bibinfo{pages}{068001}.
\newblock \DOIprefix\doi{10.1088/0031-8949/90/6/068001},
  \href{http://arxiv.org/abs/1501.03101}{\tt arXiv:1501.03101}.
%Type = Article
\bibitem[{{Przylibski} et~al.(2005){Przylibski}, {Zago{\.z}d{\.z}on}, {Kryza}
  and {Pilski}}]{Prz05}
\bibinfo{author}{{Przylibski}, T.A.}, \bibinfo{author}{{Zago{\.z}d{\.z}on},
  Pawe{\l}, P.}, \bibinfo{author}{{Kryza}, R.}, \bibinfo{author}{{Pilski},
  A.S.}, \bibinfo{year}{2005}.
\newblock \bibinfo{title}{{The Zak{\l}odzie enstatite meteorite: Mineralogy,
  petrology, origin, and classification}}.
\newblock \bibinfo{journal}{Meteoritics and Planetary Science Supplement}
  \bibinfo{volume}{40}, \bibinfo{pages}{185--200}.
%Type = Article
\bibitem[{Reichenberg(1953)}]{Rei53}
\bibinfo{author}{Reichenberg, D.}, \bibinfo{year}{1953}.
\newblock \bibinfo{title}{Properties of ion-exchange resins in relation to
  their structure. iii. kinetics of exchange}.
\newblock \bibinfo{journal}{Journal of the American Chemical Society}
  \bibinfo{volume}{75}, \bibinfo{pages}{589--597}.
\newblock \DOIprefix\doi{10.1021/ja01099a022}.
%Type = Article
\bibitem[{{Renne} et~al.(2011){Renne}, {Balco}, {Ludwig}, {Mundil} and
  {Min}}]{Ren11}
\bibinfo{author}{{Renne}, P.R.}, \bibinfo{author}{{Balco}, G.},
  \bibinfo{author}{{Ludwig}, K.R.}, \bibinfo{author}{{Mundil}, R.},
  \bibinfo{author}{{Min}, K.}, \bibinfo{year}{2011}.
\newblock \bibinfo{title}{{Response to the comment by W.H. Schwarz et al. on
  ``Joint determination of $^{40}$K decay constants and $^{40}$Ar
  $^{{\ensuremath{*}}}$/ $^{40}$K for the Fish Canyon sanidine standard, and
  improved accuracy for $^{40}$Ar/ $^{39}$Ar geochronology'' by P.R. Renne et
  al. (2010)}}.
\newblock \bibinfo{journal}{\gca} \bibinfo{volume}{75},
  \bibinfo{pages}{5097--5100}.
\newblock \DOIprefix\doi{10.1016/j.gca.2011.06.021}.
%Type = Article
\bibitem[{{Rubin} et~al.(2009){Rubin}, {Griset}, {Choi} and {Wasson}}]{Rub09}
\bibinfo{author}{{Rubin}, A.E.}, \bibinfo{author}{{Griset}, C.D.},
  \bibinfo{author}{{Choi}, B.G.}, \bibinfo{author}{{Wasson}, J.T.},
  \bibinfo{year}{2009}.
\newblock \bibinfo{title}{{Clastic matrix in EH3 chondrites}}.
\newblock \bibinfo{journal}{Meteoritics and Planetary Science}
  \bibinfo{volume}{44}, \bibinfo{pages}{589--601}.
\newblock \DOIprefix\doi{10.1111/j.1945-5100.2009.tb00754.x}.
%Type = Article
\bibitem[{{Rubin} et~al.(1997){Rubin}, {Scott} and {Keil}}]{Rub97}
\bibinfo{author}{{Rubin}, A.E.}, \bibinfo{author}{{Scott}, E.R.D.},
  \bibinfo{author}{{Keil}, K.}, \bibinfo{year}{1997}.
\newblock \bibinfo{title}{{Shock metamorphism of enstatite chondrites}}.
\newblock \bibinfo{journal}{\gca} \bibinfo{volume}{61},
  \bibinfo{pages}{847--858}.
\newblock \DOIprefix\doi{10.1016/S0016-7037(96)00364-X}.
%Type = Article
\bibitem[{{Rudee} and {Herndon}(1981)}]{Rud81}
\bibinfo{author}{{Rudee}, M.L.}, \bibinfo{author}{{Herndon}, J.M.},
  \bibinfo{year}{1981}.
\newblock \bibinfo{title}{{Thermal History of the Abee Enstatite Chondrite II;
  Thermal Measurements and Heat Flow Calculations}}.
\newblock \bibinfo{journal}{Meteoritics} \bibinfo{volume}{16},
  \bibinfo{pages}{139}.
\newblock \DOIprefix\doi{10.1111/j.1945-5100.1981.tb00538.x}.
%Type = Article
\bibitem[{{Sahijpal} and {Gupta}(2011)}]{Sah11}
\bibinfo{author}{{Sahijpal}, S.}, \bibinfo{author}{{Gupta}, G.},
  \bibinfo{year}{2011}.
\newblock \bibinfo{title}{{Did the carbonaceous chondrites evolve in the
  crustal regions of partially differentiated asteroids?}}
\newblock \bibinfo{journal}{Journal of Geophysical Research (Planets)}
  \bibinfo{volume}{116}, \bibinfo{pages}{E06004}.
\newblock \DOIprefix\doi{10.1029/2010JE003757}.
%Type = Article
\bibitem[{{Sahijpal} et~al.(2007){Sahijpal}, {Soni} and {Gupta}}]{Sah07}
\bibinfo{author}{{Sahijpal}, S.}, \bibinfo{author}{{Soni}, P.},
  \bibinfo{author}{{Gupta}, G.}, \bibinfo{year}{2007}.
\newblock \bibinfo{title}{{Numerical simulations of the differentiation of
  accreting planetesimals with 26Al and 60Fe as the heat sources}}.
\newblock \bibinfo{journal}{Meteoritics and Planetary Science}
  \bibinfo{volume}{42}, \bibinfo{pages}{1529--1548}.
\newblock \DOIprefix\doi{10.1111/j.1945-5100.2007.tb00589.x}.
%Type = Article
\bibitem[{{Schwarz} et~al.(2011){Schwarz}, {Kossert}, {Trieloff} and
  {Hopp}}]{Sch11}
\bibinfo{author}{{Schwarz}, W.H.}, \bibinfo{author}{{Kossert}, K.},
  \bibinfo{author}{{Trieloff}, M.}, \bibinfo{author}{{Hopp}, J.},
  \bibinfo{year}{2011}.
\newblock \bibinfo{title}{{Comment on the ``Joint determination of $^{40}$K
  decay constants and $^{40}$Ar $^{{\ensuremath{*}}}$/ $^{40}$K for the Fish
  Canyon sanidine standard, and improved accuracy for $^{40}$Ar/ $^{39}$Ar
  geochronology'' by Paul R. Renne et al. (2010)}}.
\newblock \bibinfo{journal}{\gca} \bibinfo{volume}{75},
  \bibinfo{pages}{5094--5096}.
\newblock \DOIprefix\doi{10.1016/j.gca.2011.06.022}.
%Type = Article
\bibitem[{{Shukolyukov} and {Lugmair}(2004)}]{Shu04}
\bibinfo{author}{{Shukolyukov}, A.}, \bibinfo{author}{{Lugmair}, G.W.},
  \bibinfo{year}{2004}.
\newblock \bibinfo{title}{{Manganese-chromium isotope systematics of enstatite
  meteorites}}.
\newblock \bibinfo{journal}{\gca} \bibinfo{volume}{68},
  \bibinfo{pages}{2875--2888}.
\newblock \DOIprefix\doi{10.1016/j.gca.2004.01.008}.
%Type = Article
\bibitem[{{Soini} et~al.(2020){Soini}, {Kukkonen}, {Kohout} and
  {Luttinen}}]{Soi20}
\bibinfo{author}{{Soini}, A.J.}, \bibinfo{author}{{Kukkonen}, I.T.},
  \bibinfo{author}{{Kohout}, T.}, \bibinfo{author}{{Luttinen}, A.},
  \bibinfo{year}{2020}.
\newblock \bibinfo{title}{{Thermal and porosity properties of meteorites: A
  compilation of published data and new measurements}}.
\newblock \bibinfo{journal}{Meteoritics and Planetary Science}
  \bibinfo{volume}{55}, \bibinfo{pages}{402--425}.
\newblock \DOIprefix\doi{10.1111/maps.13441}.
%Type = Article
\bibitem[{{Stoeffler} et~al.(1991){Stoeffler}, {Keil} and {Scott}}]{Sto91}
\bibinfo{author}{{Stoeffler}, D.}, \bibinfo{author}{{Keil}, K.},
  \bibinfo{author}{{Scott}, E.R.D.}, \bibinfo{year}{1991}.
\newblock \bibinfo{title}{{Shock metamorphism of ordinary chondrites}}.
\newblock \bibinfo{journal}{\gca} \bibinfo{volume}{55},
  \bibinfo{pages}{3845--3867}.
\newblock \DOIprefix\doi{10.1016/0016-7037(91)90078-J}.
%Type = Inproceedings
\bibitem[{{Sugiura} and {Fujiya}(2008)}]{Sug08}
\bibinfo{author}{{Sugiura}, N.}, \bibinfo{author}{{Fujiya}, W.},
  \bibinfo{year}{2008}.
\newblock \bibinfo{title}{{Al-Mg Age of the Zaklodzie Enstatite Meteorite}},
  in: \bibinfo{booktitle}{Lunar and Planetary Science Conference}, p.
  \bibinfo{pages}{1503}.
%Type = Article
\bibitem[{{Tang} and {Dauphas}(2012)}]{Tan12}
\bibinfo{author}{{Tang}, H.}, \bibinfo{author}{{Dauphas}, N.},
  \bibinfo{year}{2012}.
\newblock \bibinfo{title}{{Abundance, distribution, and origin of $^{60}$Fe in
  the solar protoplanetary disk}}.
\newblock \bibinfo{journal}{Earth and Planetary Science Letters}
  \bibinfo{volume}{359}, \bibinfo{pages}{248--263}.
\newblock \DOIprefix\doi{10.1016/j.epsl.2012.10.011},
  \href{http://arxiv.org/abs/1212.1490}{\tt arXiv:1212.1490}.
%Type = Article
\bibitem[{{Torigoye} and {Shima}(1993)}]{Tor93}
\bibinfo{author}{{Torigoye}, N.}, \bibinfo{author}{{Shima}, M.},
  \bibinfo{year}{1993}.
\newblock \bibinfo{title}{{Evidence for a Late Thermal Event of Unequilibrated
  Enstatite Chondrites: A Rb-Sr Study of Qingzhen and Yamato 6901 (EH3) and
  Khairpur (EL6)}}.
\newblock \bibinfo{journal}{Meteoritics} \bibinfo{volume}{28},
  \bibinfo{pages}{515}.
\newblock \DOIprefix\doi{10.1111/j.1945-5100.1993.tb00275.x}.
%Type = Article
\bibitem[{{Trieloff}(2009)}]{Tri09}
\bibinfo{author}{{Trieloff}, M.}, \bibinfo{year}{2009}.
\newblock \bibinfo{title}{{Chronology of the Solar System}}.
\newblock \bibinfo{journal}{Landolt B{\"o}rnstein} \bibinfo{volume}{4B},
  \bibinfo{pages}{771}.
\newblock \DOIprefix\doi{10.1007/978-3-540-88055-4\_35}.
%Type = Article
\bibitem[{{Trieloff} et~al.(1998){Trieloff}, {Deutsch} and
  {Jessberger}}]{Tri98}
\bibinfo{author}{{Trieloff}, M.}, \bibinfo{author}{{Deutsch}, A.},
  \bibinfo{author}{{Jessberger}, E.K.}, \bibinfo{year}{1998}.
\newblock \bibinfo{title}{{The age of the Kara impact structure, Russia}}.
\newblock \bibinfo{journal}{Meteoritics and Planetary Science}
  \bibinfo{volume}{33}, \bibinfo{pages}{361--372}.
\newblock \DOIprefix\doi{10.1111/j.1945-5100.1998.tb01640.x}.
%Type = Article
\bibitem[{{Trieloff} et~al.(2003){Trieloff}, {Jessberger}, {Herrwerth}, {Hopp},
  {Fi{\'e}ni}, {Gh{\'e}lis}, {Bourot-Denise} and {Pellas}}]{Tri03}
\bibinfo{author}{{Trieloff}, M.}, \bibinfo{author}{{Jessberger}, E.K.},
  \bibinfo{author}{{Herrwerth}, I.}, \bibinfo{author}{{Hopp}, J.},
  \bibinfo{author}{{Fi{\'e}ni}, C.}, \bibinfo{author}{{Gh{\'e}lis}, M.},
  \bibinfo{author}{{Bourot-Denise}, M.}, \bibinfo{author}{{Pellas}, P.},
  \bibinfo{year}{2003}.
\newblock \bibinfo{title}{{Structure and thermal history of the H-chondrite
  parent asteroid revealed by thermochronometry}}.
\newblock \bibinfo{journal}{\nat} \bibinfo{volume}{422},
  \bibinfo{pages}{502--506}.
\newblock \DOIprefix\doi{10.1038/nature01499}.
%Type = Inbook
\bibitem[{{Trieloff} and {Palme}(2006)}]{Tri06}
\bibinfo{author}{{Trieloff}, M.}, \bibinfo{author}{{Palme}, H.},
  \bibinfo{year}{2006}.
\newblock \bibinfo{title}{{The origin of solids in the early Solar System}}.
  \bibinfo{publisher}{Cambridge University Press},
  \bibinfo{address}{Cambridge}.
\newblock p.~\bibinfo{pages}{64}.
%Type = Article
\bibitem[{Trieloff et~al.(1994)Trieloff, Reimold, Kunz, Boer and
  Jessberger}]{Tri94}
\bibinfo{author}{Trieloff, M.}, \bibinfo{author}{Reimold, W.},
  \bibinfo{author}{Kunz, J.}, \bibinfo{author}{Boer, R.},
  \bibinfo{author}{Jessberger, E.}, \bibinfo{year}{1994}.
\newblock \bibinfo{title}{40ar-39ar thermochronology of pseudotachylite at the
  ventersdorp contact reef, witwatersrand basin}.
\newblock \bibinfo{journal}{South African Journal of Geology}
  \bibinfo{volume}{97}, \bibinfo{pages}{365--384}.
\newblock \URLprefix
  \url{https://journals.co.za/content/sajg/97/3/AJA10120750_876}.
%Type = Article
\bibitem[{{Turner}(1979)}]{Tur79}
\bibinfo{author}{{Turner}, G.}, \bibinfo{year}{1979}.
\newblock \bibinfo{title}{{A Monte Carlo Model for the Production of Meteorites
  with Implications for Gas Retention Ages}}.
\newblock \bibinfo{journal}{Meteoritics} \bibinfo{volume}{14},
  \bibinfo{pages}{550}.
%Type = Inproceedings
\bibitem[{{Uribe} et~al.(2016a){Uribe}, {Izawa}, {McCausland} and
  {Flemming}}]{Uri16a}
\bibinfo{author}{{Uribe}, D.D.}, \bibinfo{author}{{Izawa}, M.R.M.},
  \bibinfo{author}{{McCausland}, P.J.A.}, \bibinfo{author}{{Flemming}, R.L.},
  \bibinfo{year}{2016}a.
\newblock \bibinfo{title}{{Mineralogy, Petrology, and Mineral Chemistry of
  Northwest Africa 8173: An Anomalous Enstatite Achondrite with Evidence for
  High-Temperature Silicate Sulphidation}}, in: \bibinfo{booktitle}{Lunar and
  Planetary Science Conference}, p. \bibinfo{pages}{2797}.
%Type = Inproceedings
\bibitem[{{Uribe} et~al.(2016b){Uribe}, {McCausland} and {Izawa}}]{Uri16b}
\bibinfo{author}{{Uribe}, D.D.}, \bibinfo{author}{{McCausland}, P.J.A.},
  \bibinfo{author}{{Izawa}, M.R.M.}, \bibinfo{year}{2016}b.
\newblock \bibinfo{title}{{A Comparative Study of the Zaklodzie and Northwest
  Africa 4301 Anomalous Enstatite Achondrites}}, in: \bibinfo{booktitle}{Lunar
  and Planetary Science Conference}, p. \bibinfo{pages}{3071}.
%Type = Article
\bibitem[{{Van Schmus}(1969)}]{vSc69}
\bibinfo{author}{{Van Schmus}, W.R.}, \bibinfo{year}{1969}.
\newblock \bibinfo{title}{{The mineralogy and petrology of chondritic
  meteorites}}.
\newblock \bibinfo{journal}{Earth Science Reviews} \bibinfo{volume}{5},
  \bibinfo{pages}{145--184}.
\newblock \DOIprefix\doi{10.1016/0012-8252(69)90076-2}.
%Type = Article
\bibitem[{{Visser} and {Ormel}(2016)}]{Vis16}
\bibinfo{author}{{Visser}, R.G.}, \bibinfo{author}{{Ormel}, C.W.},
  \bibinfo{year}{2016}.
\newblock \bibinfo{title}{{On the growth of pebble-accreting planetesimals}}.
\newblock \bibinfo{journal}{\aap} \bibinfo{volume}{586}, \bibinfo{pages}{A66}.
\newblock \DOIprefix\doi{10.1051/0004-6361/201527361},
  \href{http://arxiv.org/abs/1511.03903}{\tt arXiv:1511.03903}.
%Type = Article
\bibitem[{{Weidenschilling}(1977)}]{Wei77}
\bibinfo{author}{{Weidenschilling}, S.J.}, \bibinfo{year}{1977}.
\newblock \bibinfo{title}{{Aerodynamics of solid bodies in the solar nebula.}}
\newblock \bibinfo{journal}{\mnras} \bibinfo{volume}{180},
  \bibinfo{pages}{57--70}.
\newblock \DOIprefix\doi{10.1093/mnras/180.2.57}.
%Type = Article
\bibitem[{{Weisberg} and {Kimura}(2012)}]{Wei12}
\bibinfo{author}{{Weisberg}, M.K.}, \bibinfo{author}{{Kimura}, M.},
  \bibinfo{year}{2012}.
\newblock \bibinfo{title}{{The unequilibrated enstatite chondrites}}.
\newblock \bibinfo{journal}{Chemie der Erde / Geochemistry}
  \bibinfo{volume}{72}, \bibinfo{pages}{101--115}.
\newblock \DOIprefix\doi{10.1016/j.chemer.2012.04.003}.
%Type = Inproceedings
\bibitem[{{Weisberg} et~al.(2014){Weisberg}, {Zolensky}, {Kimura} and
  {Ebel}}]{Wei14}
\bibinfo{author}{{Weisberg}, M.K.}, \bibinfo{author}{{Zolensky}, M.E.},
  \bibinfo{author}{{Kimura}, M.}, \bibinfo{author}{{Ebel}, D.S.},
  \bibinfo{year}{2014}.
\newblock \bibinfo{title}{{Primitive Fine-Grained Matrix in the Unequilibrated
  Enstatite Chondrites}}, in: \bibinfo{booktitle}{45th Lunar and Planetary
  Science Conference}, p. \bibinfo{pages}{1551}.
%Type = Article
\bibitem[{{Yomogida} and {Matsui}(1983)}]{Yom83}
\bibinfo{author}{{Yomogida}, K.}, \bibinfo{author}{{Matsui}, T.},
  \bibinfo{year}{1983}.
\newblock \bibinfo{title}{{Physical properties of ordinary chondrites and their
  implications.}}
\newblock \bibinfo{journal}{Meteoritics} \bibinfo{volume}{18},
  \bibinfo{pages}{430--431}.
%Type = Article
\bibitem[{{Youdin} and {Goodman}(2005)}]{You05}
\bibinfo{author}{{Youdin}, A.N.}, \bibinfo{author}{{Goodman}, J.},
  \bibinfo{year}{2005}.
\newblock \bibinfo{title}{{Streaming Instabilities in Protoplanetary Disks}}.
\newblock \bibinfo{journal}{\apj} \bibinfo{volume}{620},
  \bibinfo{pages}{459--469}.
\newblock \DOIprefix\doi{10.1086/426895}.
%Type = Article
\bibitem[{{Zellner} et~al.(1977){Zellner}, {Leake}, {Morrison} and
  {Williams}}]{Zel77}
\bibinfo{author}{{Zellner}, B.}, \bibinfo{author}{{Leake}, M.},
  \bibinfo{author}{{Morrison}, D.}, \bibinfo{author}{{Williams}, J.G.},
  \bibinfo{year}{1977}.
\newblock \bibinfo{title}{{The E asteroids and the origin of the enstatite
  achondrites}}.
\newblock \bibinfo{journal}{\gca} \bibinfo{volume}{41},
  \bibinfo{pages}{1759--1767}.
\newblock \DOIprefix\doi{10.1016/0016-7037(77)90208-3}.
%Type = Article
\bibitem[{{Zhang} et~al.(2020){Zhang}, {Mei} and {Song}}]{Zha20}
\bibinfo{author}{{Zhang}, G.}, \bibinfo{author}{{Mei}, S.},
  \bibinfo{author}{{Song}, M.}, \bibinfo{year}{2020}.
\newblock \bibinfo{title}{{Effect of Water on the Dislocation Creep of
  Enstatite Aggregates at 300 MPa}}.
\newblock \bibinfo{journal}{Geophysical Research Letters} \bibinfo{volume}{47},
  \bibinfo{pages}{e85895}.
\newblock \DOIprefix\doi{10.1029/2019GL085895}.
%Type = Article
\bibitem[{{Zhang} et~al.(1996){Zhang}, {Huang}, {Schneider}, {Benoit},
  {DeHart}, {Lofgren} and {Sears}}]{Zha96b}
\bibinfo{author}{{Zhang}, Y.}, \bibinfo{author}{{Huang}, S.},
  \bibinfo{author}{{Schneider}, D.}, \bibinfo{author}{{Benoit}, P.H.},
  \bibinfo{author}{{DeHart}, J.M.}, \bibinfo{author}{{Lofgren}, G.E.},
  \bibinfo{author}{{Sears}, D.W.G.}, \bibinfo{year}{1996}.
\newblock \bibinfo{title}{{Pyroxene structures, cathodoluminescence and the
  thermal history of the enstatite chondrites}}.
\newblock \bibinfo{journal}{\maps} \bibinfo{volume}{31},
  \bibinfo{pages}{87--96}.
\newblock \DOIprefix\doi{10.1111/j.1945-5100.1996.tb02058.x}.
%Type = Article
\bibitem[{{Zhang} and {Sears}(1996)}]{Zha96}
\bibinfo{author}{{Zhang}, Y.}, \bibinfo{author}{{Sears}, D.W.G.},
  \bibinfo{year}{1996}.
\newblock \bibinfo{title}{{The thermometry of enstatite chondrites: A brief
  review and update}}.
\newblock \bibinfo{journal}{Meteoritics and Planetary Science}
  \bibinfo{volume}{31}, \bibinfo{pages}{647--655}.
\newblock \DOIprefix\doi{10.1111/j.1945-5100.1996.tb02038.x}.
%Type = Article
\bibitem[{{Zhu} et~al.(2020){Zhu}, {Moynier}, {Schiller} and
  {Bizzarro}}]{Zhu20}
\bibinfo{author}{{Zhu}, K.}, \bibinfo{author}{{Moynier}, F.},
  \bibinfo{author}{{Schiller}, M.}, \bibinfo{author}{{Bizzarro}, M.},
  \bibinfo{year}{2020}.
\newblock \bibinfo{title}{{Dating and Tracing the Origin of Enstatite Chondrite
  Chondrules with Cr Isotopes}}.
\newblock \bibinfo{journal}{\apjl} \bibinfo{volume}{894}, \bibinfo{pages}{L26}.
\newblock \DOIprefix\doi{10.3847/2041-8213/ab8dca}.
%Type = Article
\bibitem[{{Zipfel} et~al.(2010){Zipfel}, {Bischoff}, {Schultz}, {Spettel},
  {Dreibus}, {Sch{\"o}nbeck} and {Palme}}]{Zip10}
\bibinfo{author}{{Zipfel}, J.}, \bibinfo{author}{{Bischoff}, A.},
  \bibinfo{author}{{Schultz}, L.}, \bibinfo{author}{{Spettel}, B.},
  \bibinfo{author}{{Dreibus}, G.}, \bibinfo{author}{{Sch{\"o}nbeck}, T.},
  \bibinfo{author}{{Palme}, H.}, \bibinfo{year}{2010}.
\newblock \bibinfo{title}{{Mineralogy, chemistry, and irradiation record of
  Neuschwanstein (EL6) chondrite}}.
\newblock \bibinfo{journal}{Meteoritics and Planetary Science}
  \bibinfo{volume}{45}, \bibinfo{pages}{1488--1501}.
\newblock \DOIprefix\doi{10.1111/j.1945-5100.2010.01120.x}.

\end{thebibliography}

\end{document}